\documentclass[prc]{revtex4}
\usepackage{graphicx}
\begin{document}

\newcommand\ie {{\it i.e.}}
\newcommand\eg {{\it e.g.}}
\newcommand\etc{{\it etc.}}
\newcommand\cf {{\it cf.}}
\newcommand\etal {{\it et al.}}
\newcommand{\be}{\begin{eqnarray}}
\newcommand{\ee}{\end{eqnarray}}
\newcommand{\jp}{$ J/ \psi $}
\newcommand{\pp}{$ \psi^{ \prime} $}
\newcommand{\ppp}{$ \psi^{ \prime \prime } $}
\newcommand{\dd}[2]{$ #1 \overline #2 $}
\newcommand\noi {\noindent}

\title{Limits on Intrinsic Charm Production from the SeaQuest Experiment}

\author{R. Vogt}
\affiliation
{Nuclear and Chemical Sciences Division,
Lawrence Livermore National Laboratory, Livermore, CA 94551,
USA}
\affiliation
    {Department of Physics and Astronomy,
University of California, Davis, CA 95616,
USA}

\begin{abstract}
  {\bf Background:} A nonperturbative charm production contribution, known as
  intrinsic charm, has long been speculated but has never been satisfactorily
  proven.  The
  SeaQuest experiment at FNAL is in an ideal kinematic region to provide
  evidence of $J/\psi$ production by intrinsic charm.
  {\bf Purpose:} $J/\psi$ production in the SeaQuest kinematics is calculated
  with a combination of perturbative QCD and intrinsic charm to see whether the
  SeaQuest data can put limits on an intrinsic charm contribution.
  {\bf Methods:} $J/\psi$ production in perturbative QCD is calculated to
  next-to-leading order in the cross section.   Cold nuclear matter effects
  include nuclear modification of the parton densities, absorption by nucleons,
  and $p_T$ broadening by multiple scattering.  The $J/\psi$ contribution from
  intrinsic charm is calculated assuming production from a
  $|uud c \overline c \rangle$ Fock state.
  {\bf Results:}  The nuclear modification factor, $R_{pA}$, is calculated as a
  function of $x_F$ and $p_T$ for $p+$C, $p+$Fe, and $p+$W interactions
  relative to $p+$d.
  {\bf Conclusions:}  The SeaQuest kinematic acceptance is ideal for testing
  the limits on intrinsic charm in the proton.
\end{abstract}
\maketitle

\section{Introduction}

Production of $J/\psi$ has long been studied in a variety of environments, from
more elementary collisions such as $p+p$ and $p+\overline p$, to protons on
nuclei, $p+A$, and in nucleus-nucleus collisions, $A+A$.  While many of the
earlier studies of $J/\psi$ production in $p+A$ collisions were made in
fixed-target configurations \cite{NA3,E537,E772,E789,na50,e866,herab1,herab2},
more recently production has predominantly been studied at hadron and nuclear
colliders, in particular at
the Relativistic Heavy-Ion Collider at Brookhaven \cite{PHENIX,STAR}
and the Large Hadron Collider at CERN \cite{ALICE,LHCb,CMS,ATLAS}.  However, in
the last several years, $J/\psi$ production has been studied by the SeaQuest
experiment at Fermilab \cite{SeaQuest}
in a fixed-target setup and at a lower incident proton
energy than previous experiments, allowing unprecedented forward coverage
probing partons carrying a large fraction, $x_1$, of the incident proton
momentum.

This large $x_1$ coverage is of interest because it is ideal for setting
limits on any $c \overline c$ component of the proton wavefunction, proposed
by Brodsky and collaborators \cite{intc1,intc2,BandH}.  There have been some
tantalizing hints \cite{NA3,EMC,ISR} but no concrete confirmation so far.
Fixed-target data usually do not cover the far forward $x_F$ region or have
done so with insufficient statistics to make any definitive statement.
Production at collider energies is ineffective for such studies because the
collider kinematics put large $x$ partons from the projectiles outside the
detector coverage \cite{RV_SJB_19}.
Several additional new fixed-target experiments have been
proposed \cite{NA60+,AFTER} or have taken data \cite{SMOG}, including using
the LHC for fixed-target studies \cite{AFTER,SMOG}, but all are higher energy
than SeaQuest.

Earlier studies of $J/\psi$ production at fixed-target experiments
parameterized the nuclear dependence of $J/\psi$ production as a power law 
\cite{NA3,E537,E772,E789,na50,e866,herab1,herab2},
\be
\sigma_{pA} = \sigma_{pN} A^\alpha \, \, , \label{alfint}
\ee where $A$ is the target mass and $\sigma_{pA}$ and $\sigma_{pN}$ are the
production cross sections in proton-nucleus and
proton-nucleon interactions respectively.  
If $J/\psi$ production is unaffected by the presence of the nucleus,
$\sigma_{pA}$ would grow linearly with $A$ and $\alpha = 1$.  
However, a less than linear $A$ dependence has been observed
\cite{NA3,E537,E772,E789,na50,e866,herab1,herab2}.  Typical values of 
$\alpha$ were between 0.9 and 1.  
This $A$ dependence was previously used
to determine an effective nuclear absorption cross section
\cite{na50,rvrev,klns}
under the assumption that the deviation of $\alpha$ from unity was solely due
to $J/\psi$ dissociation by nucleons.  However, more detailed studies of
$\alpha$ as a function of the longitudinal momentum fraction, $x_F$, and
transverse momentum, $p_T$, revealed a more complex picture.
Indeed, it has long been known 
that $\alpha$ decreases as a function of $x_F$
\cite{NA3,E537,E772,herab2} and increases with $p_T$ \cite{E772,e866,herab2},
neither of
which can be accounted for by nuclear absorption alone.  A number of cold
nuclear matter effects can contribute to this dependence.  These include
nuclear modifications of the parton distributions in nucleons in a nucleus
relative to those of a free proton \cite{Arn}, and
energy loss or transverse momentum broadening due to multiple scattering as
the projectile transits the nucleus \cite{GM,BH,Arleo}, in addition to
absorption.  It has however, proven difficult for most model calculations
to describe $J/\psi$ production on nuclear targets as a function of $x_F$
and $p_T$ with these cold nuclear matter effects alone.

The NA3 Collaboration \cite{NA3} proposed dividing $J/\psi$ production into
two components that they referred to as hard and diffractive for the nuclear
volume-like and surface-like dependencies respectively,
\be  \frac{d\sigma_{pA}}{dx_F}  & = &
A^{\alpha^\prime} \frac{d\sigma_h}{dx_F} + A^\beta \frac{d\sigma_d}{dx_F} \, \,
. \label{twocom} \ee 
The hard component is the perturbative QCD contribution discussed so far
while the so-called
diffractive component has been attributed to intrinsic charm,
$c \overline c$ pairs in the proton wavefunction
\cite{intc1,intc2,BandH}. 
Since the charm quark mass is large, these intrinsic heavy quark pairs 
carry a significant fraction
of the longitudinal momentum and contribute at large $x_F$  
whereas perturbative charm production decreases strongly with $x_F$.  
The light spectator quarks in the intrinsic \dd{c}{c} state 
interact on the nuclear surface, leading to an approximate $A^{2/3}$ dependence
\cite{BandH}.  The NA3 Collaboration found $\alpha^\prime =0.97$ and
$\beta = 0.71$ for proton projectiles \cite{NA3}. This value of $\beta$ is
approximately consistent with a nuclear dependence due to surface interactions
relative to interactions with the nuclear volume, as in hard scattering
\cite{BandH}.  Such an approach was employed to compare to the NA3
measurements \cite{VBH1} for incident protons at 200~GeV and other
fixed-target experiments at higher energies \cite{RV_e866}.

The SeaQuest experiment at FNAL, with a 120~GeV incident proton beam,
is in an ideal kinematic regime to test intrinsic charm production.  Its
$J/\psi$ acceptance is in the region $0.4 < x_F < 0.95$ and $p_T < 2.3$~GeV.
They have measured the $x_F$ and $p_T$ dependence of $J/\psi$ production in
$p+p$, $p+{\rm d}$, $p+{\rm C}$, $p+ {\rm Fe}$ and $p+ {\rm W}$
interactions \cite{Ayuso}.  A similar two-component
model is employed to make predictions for these measurements,
at energies and with kinematic
acceptance favorable to intrinsic charm, in this work.  
The results are expressed through the nuclear modification, $R_{pA}$, 
factor, the ratio of the per nucleon cross section in $p+A$ collisions relative
to $p+{\rm d}$
interactions at the same energy, instead of $\alpha$, as previously
employed.

Here $J/\psi$ production in perturbative QCD is presented in Sec.~\ref{pQCD} and
the cold nuclear matter effects included in the calculation are introduced.
The $x_F$ and $p_T$ distributions from intrinsic charm is presented in
Sec.~\ref{ICcomp}, along with a discussion of its nuclear dependence.
Section~\ref{model_comp} presents the results for the modification of $J/\psi$
production in nuclear targets at SeaQuest, both for perturbative QCD effects
alone and including the intrinsic charm contribution.
The conclusions are presented in Sec.~\ref{conclusions}.

\section{$J/\psi$ Production and Cold Nuclear Matter Effects in Perturbative
  QCD}
\label{pQCD}

This section describes $J/\psi$ production in perturbative QCD and describes
how cold nuclear matter effects are implemented in this work.

The $J/\psi$ production mechanism remains an unsettled question, with a number
of approaches having been introduced \cite{HPC,NRQCD,ICEM}.  In the calculations
presented here, the Color Evaporation Model \cite{HPC} is employed.  This model,
together with the Improved Color Evaporation Model \cite{ICEM}, can describe
the $x_F$ and $p_T$ distributions of $J/\psi$ production, including at low
$p_T$ where other approaches have some difficulties and may require a $p_T$
cut \cite{QWG_rev}.  The CEM calculations are described in Sec.~\ref{CEM}.  The
remainder of this section discusses nuclear modifications of the parton
densities, nPDF effects, Sec.~\ref{shad};
absorption by nucleons, Sec.~\ref{absorption}; and transverse momentum
broadening, Sec.~\ref{kTkick}; are implemented in this work.

\subsection{Charmonium Production in the Color Evaporation Model}
\label{CEM}

The Color
Evaporation Model \cite{HPC} is employed to calculate $J/\psi$ production.
This approach assumes that some fraction, $F_C$, of the $c \overline c$ pairs
produced in perturbative QCD with a pair mass below 
that of the $D \overline D$ pair mass threshold
will go on mass shell as a $J/\psi$,
\be
\sigma_{\rm CEM}(pp) = F_C \sum_{i,j} 
\int_{4m^2}^{4m_H^2} ds
\int dx_1 \, dx_2~ F_i^p(x_1,\mu_F^2,k_{T_1})~ F_j^p(x_2,\mu_F^2,k_{T_2})~ 
\hat\sigma_{ij}(\hat{s},\mu_F^2, \mu_R^2) \, \, , 
\label{sigCEM}
\ee
where $ij = gg, q\overline q$ or $q(\overline q)g$ and
$\hat\sigma_{ij}(\hat {s},\mu_F^2, \mu_R^2)$
is the partonic cross section for initial state $ij$ evaluated at factorization
scale $\mu_F$ and renormalization scale $\mu_R$.  (Note that the
$q(\overline q)g$ process only enters at next-to-leading order in $\alpha_s$.)
The parton densities are
written here to include intrinsic $k_T$, required to keep the pair cross section
finite as $p_T \rightarrow 0$.  They are assumed to factorize into the normal
parton densities in collinear factorization and a $k_T$-dependent function,
\be
F^p(x,\mu_F^2,k_T) = f^p(x,\mu_F^2)G_p(k_T) \, \, . \label{PDFfact}
\ee
The CT10 proton parton densities
\cite{CT10} are employed in the calculations of $f^p(x,\mu_F^2)$.

The charm quark
mass, factorization scale and renormalization scale determined from a fit to
the total $c \overline c$
cross section at NLO in Ref.~\cite{NVF} are employed here,
$(m,\mu_F/m_T, \mu_R/m_T) = (1.27 \pm 0.09 \, {\rm GeV}, 2.1^{+2.55}_{-0.85}, 1.6^{+0.11}_{-0.12})$ where $\mu_F$ is the factorization scale and $\mu_R$ is the
renormalization scale.   The scale factors, $\mu_F$ and
$\mu_R$, are defined relative to the transverse mass of the pair,
$\mu_{F,R} \propto m_T = \sqrt{m^2 + p_T^2}$ where 
the $p_T$ is the $c \overline c$ pair $p_T$, 
$p_T^2 = 0.5(p_{T_Q}^2 + p_{T_{\overline Q}}^2)$.  At LO in the CEM,
the $J/\psi$ $p_T$ is zero.  Thus, the calculated $p_T$ distribution is LO
while the $x_F$ distribution is at NLO.  

The Improved Color
Evaporation Model has recently been developed \cite{ICEM} and extended to
studies of quarkonium polarization \cite{CV1,CV2,CV3,CV4} in hadroproduction.
The change in the mass integration range and in the definition of the $p_T$ of
the quarkonium state changes the $\psi$(2S) $p_T$ distribution
relative to the $J/\psi$ at high $p_T$
and, due to the narrower integration range, increases $F_C$ for $J/\psi$
in the ICEM by
$\sim 40$\% \cite{ICEM}.  The $J/\psi$ $p_T$ distributions are compatible with
each other in the two approaches, see Ref.~\cite{CV3} for a comparison.

Calculations in the CEM were carried out at next-to-leading order (NLO)
in the heavy flavor cross section using the exclusive NLO $Q \overline Q$
production code HVQMNR \cite{MNR} after imposing a cut on the
pair invariant mass.  This exclusive calculation is applied because it can keep
track of both the $c$ and $\overline c$ quarks that constitute the $J/\psi$.
However, to keep the $p_T$ distribution finite as
$p_T \rightarrow 0$, the calculations require
augmentation by $k_T$ broadening, as described in more detail below.

Results on open heavy flavors at fixed-target energies 
indicated that some level of
transverse momentum broadening was needed to obtain agreement with the
fixed-target data after fragmentation was applied \cite{MLM1}.
Broadening has typically been modeled by including intrinsic transverse
momentum, $k_T$, smearing to the parton densities and plays the role of
low $p_T$ QCD resummation \cite{CYLO}.  

In the HVQMNR code, an intrinsic $k_T$ is added in the
final state, rather than the initial state, as was previously done for
Drell-Yan production \cite{CYLO}. 
When applied in the intial-state it multiplies the parton
distribution functions for both hadrons, 
assuming the $x$ and $k_T$ dependencies factorize in the parton distributions,
as in Eq.~(\ref{PDFfact}).  
If the $k_T$ kick is not too large, it does not matter whether
the $k_T$ dependence is added in the initial or final state.

Because the kick is employed in the final state, effectively the
factors $G_p(k_T)$ in Eq.~(\ref{PDFfact}) for each parton is replaced by
\begin{eqnarray}
  g_p(k_T) = G_p(k_{T_1})G_p(k_{T_2}) \, \, . \label{FSgauss}
\end{eqnarray}
A Gaussian is employed to define $g_p(k_T)$ \cite{MLM1}, 
\begin{eqnarray}
g_p(k_T) = \frac{1}{\pi \langle k_T^2 \rangle_p} \exp(-k_T^2/\langle k_T^2
\rangle_p) \, \, .
\label{intkt}
\end{eqnarray}
In Ref.~\cite{MLM1}, $\langle k_T^2 \rangle_p = 1$ GeV$^2$ was chosen
to describe the $p_T$ dependence of fixed-target charm production.  

The broadening is applied by first boosting the produced $c \overline c$ pair
(plus light parton at NLO)
to its rest frame
from its longitudinal center-of-mass frame.  The intrinsic transverse
momenta, $\vec k_{T 1}$ and $\vec k_{T 2}$, of the incident partons are chosen
at random with $k_{T 1}^2$ and $k_{T 2}^2$ distributed according to
Eq.~(\ref{intkt}), preserving the total kick.
Boosting back to the original frame changes the initial transverse momentum of
the $c \overline c$ pair from $\vec p_T$ to
$\vec p_T + \vec k_{T 1} + \vec k_{T 2}$.  If the $k_T$ was applied to
the incident parton distributions, instead of the final-state $c \overline c$
pair, there would be no difference at LO.  At NLO there can also be a
light parton in the final state, making the correspondence inexact.  
The difference between the two implementations should be small if
$\langle k_T^2 \rangle \leq 2$ GeV$^2$, as is the case here
\cite{MLM1}.  While the
rapidity distributions are independent of the intrinsic $k_T$, there will be
some modification of the $x_F$ distribution because
$x_F = (2m_T/\sqrt{s_{NN}})\sinh y$ and $m_T = \sqrt{p_T^2 + m^2}$.

The effect of the $k_T$ kick on the $p_T$ distribution can be expected to
decrease as $\sqrt{s}$ increases because the average $p_T$ also increases 
with energy.  However, the value of $\langle k_T^2 \rangle_p$ is assumed to
increase with $\sqrt{s}$ so that effect remains particularly
important for low $p_T$ production at higher energies.
The energy dependence of $\langle k_T^2 \rangle$ in Ref.~\cite{NVF} is
\begin{eqnarray}
  \langle k_T^2 \rangle_p = \left[ 1 + \frac{1}{n} \ln
    \left(\frac{\sqrt{s_{NN}} ({\rm GeV})}{20 \,
    {\rm GeV}} \right) \right] \, \, {\rm GeV}^2 \, \, 
\label{eq:avekt}
\end{eqnarray}
with $n = 12$ for $J/\psi$ production \cite{NVF}.  At the energy of the SeaQuest
experiment, $\sqrt{s_{NN}} = 15.4$~GeV, $\langle k_T^2 \rangle_p = 0.97$~GeV$^2$.

\begin{figure}
  \begin{center}
    \includegraphics[width=0.495\textwidth]{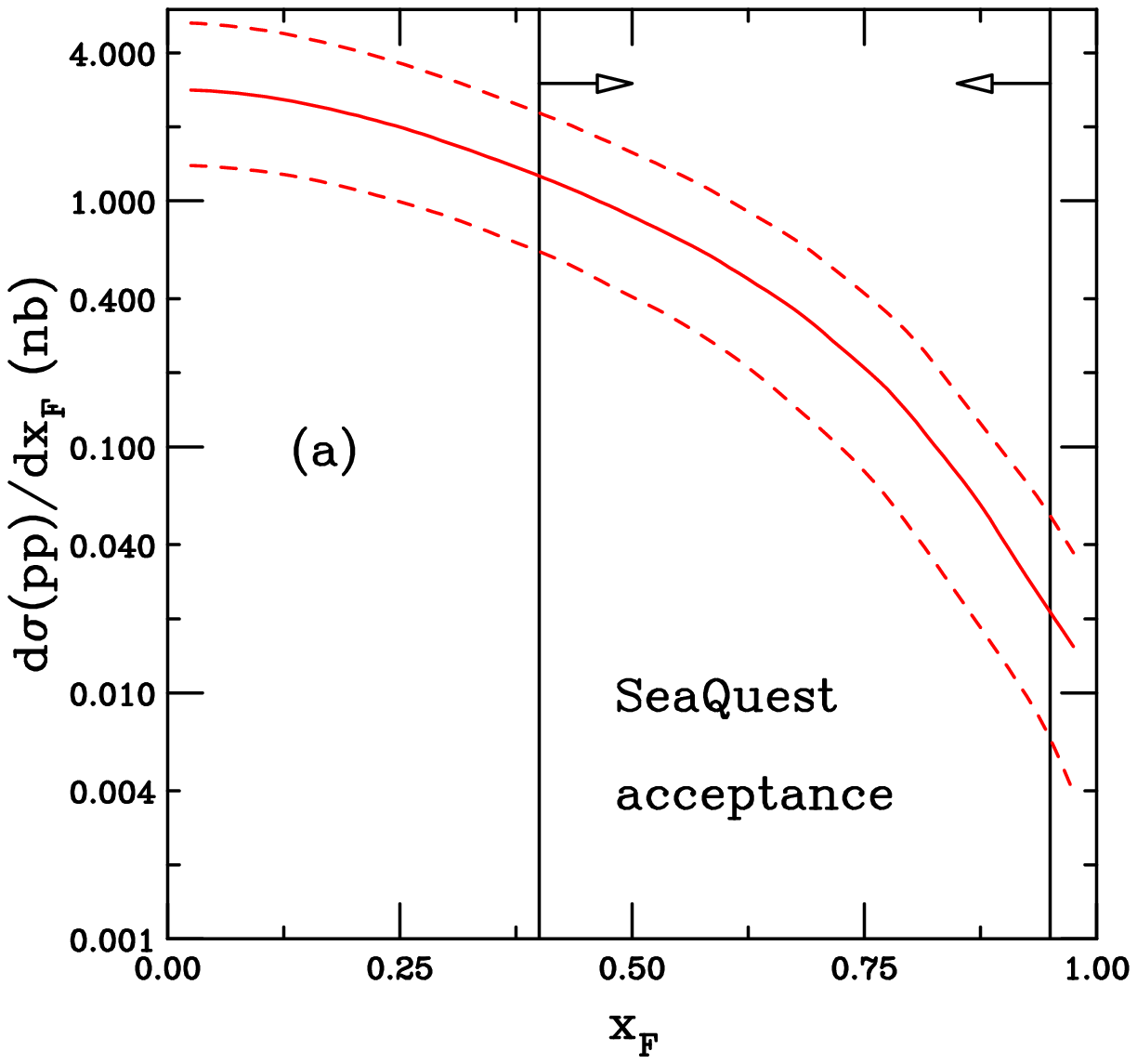}    
    \includegraphics[width=0.495\textwidth]{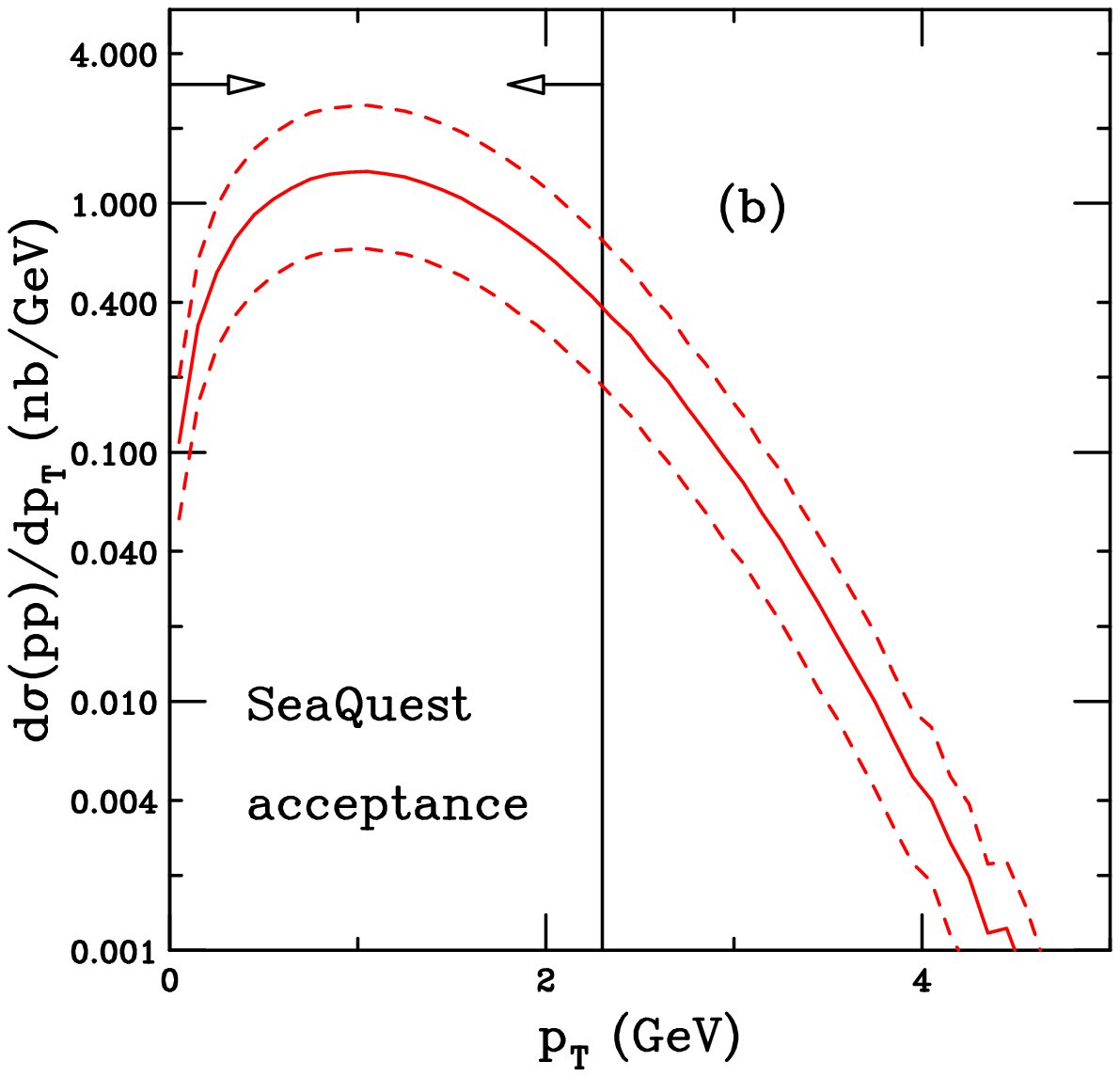}    
  \end{center}
\caption[]{(Color online) The $J/\psi$ production cross sections in the CEM
  in $p+p$ collisions at $\sqrt{s} = 15.4$~GeV as a function of $x_F$ (a) and
  $p_T$ (b), integrated over all phase space, are shown.  The solid curves show
  the central values while the dashed curves outline the upper and lower limits
  of the uncertainty band.}
\label{cem_pp}
\end{figure}

Figure~\ref{cem_pp}
shows the $x_F$ and $p_T$ distributions for $J/\psi$ production
in $p+p$ collisions in the CEM at 120 GeV.  
The mass and scale uncertainties are calculated 
using the one standard deviation uncertainties on
the quark mass and scale parameters.  If the central, upper and lower limits
of $\mu_{R,F}/m_T$ are denoted as $C$, $H$, and $L$ respectively, the seven
sets used to determine the scale uncertainty are  $\{(\mu_F/m_T,\mu_F/m_T)\}$ =
$\{$$(C,C)$, $(H,H)$, $(L,L)$, $(C,L)$, $(L,C)$, $(C,H)$, $(H,C)$$\}$.    
The uncertainty band can be obtained for the best fit sets
\cite{NVF} by
adding the uncertainties from the mass and scale variations in 
quadrature. The envelope contained by the resulting curves,
\begin{eqnarray}
\frac{d\sigma_{\rm max}}{dX} & = & \frac{d\sigma_{\rm cent}}{dX} 
+ \sqrt{\left(\frac{d\sigma_{\mu ,{\rm max}}}{dX} -
  \frac{d\sigma_{\rm cent}}{dX}\right)^2
  + \left(\frac{d\sigma_{m, {\rm max}}}{dX} -
  \frac{d\sigma_{\rm cent}}{dX}\right)^2} \, \, , \label{sigmax}
\\
\frac{d\sigma_{\rm min}}{dX} & = & \frac{d\sigma_{\rm cent}}{dX} 
- \sqrt{\left(\frac{d\sigma_{\mu ,{\rm min}}}{dX} -
  \frac{d\sigma_{\rm cent}}{dX}\right)^2
  + \left(\frac{d\sigma_{m, {\rm min}}}{dX} -
  \frac{d\sigma_{\rm cent}}{dX}\right)^2} \, \, , \label{sigmin}  
\end{eqnarray}
defines the uncertainty on the cross section.
The kinematic observables, denoted by $X$, are $x_F$ and $p_T$ in this case.
The calculation labeled ``cent'' employs the central values of
$m$, $\mu_F$ and $\mu_R$.  The calculations with subscript
$\mu$ keep the mass fixed to the central value while the scales are varied.
On the other hand, in the calculations with subscript $m$, the scales are fixed
to their central values while the mass is varied
between its upper and lower limits.  Unless otherwise noted, the
calculations employ the central values of the mass and scale
factors.

At the center of mass energy of the SeaQuest experiment, the $J/\psi$
cross section is still growing rather quickly with $\sqrt{s}$ and the mass
uncertainty is dominant.  Changing the charm quark mass within its 7\%
uncertainty can change the forward $J/\psi$ cross section (in the range
$0 < x_F < 1$) by almost a factor of
two.  On the other hand, changing the factorization scale at the relatively
large momentum fraction carried by the incident partons, gives a rather small
effect on the cross sections, less than 30\%.

\begin{figure}
  \begin{center}
    \includegraphics[width=0.495\textwidth]{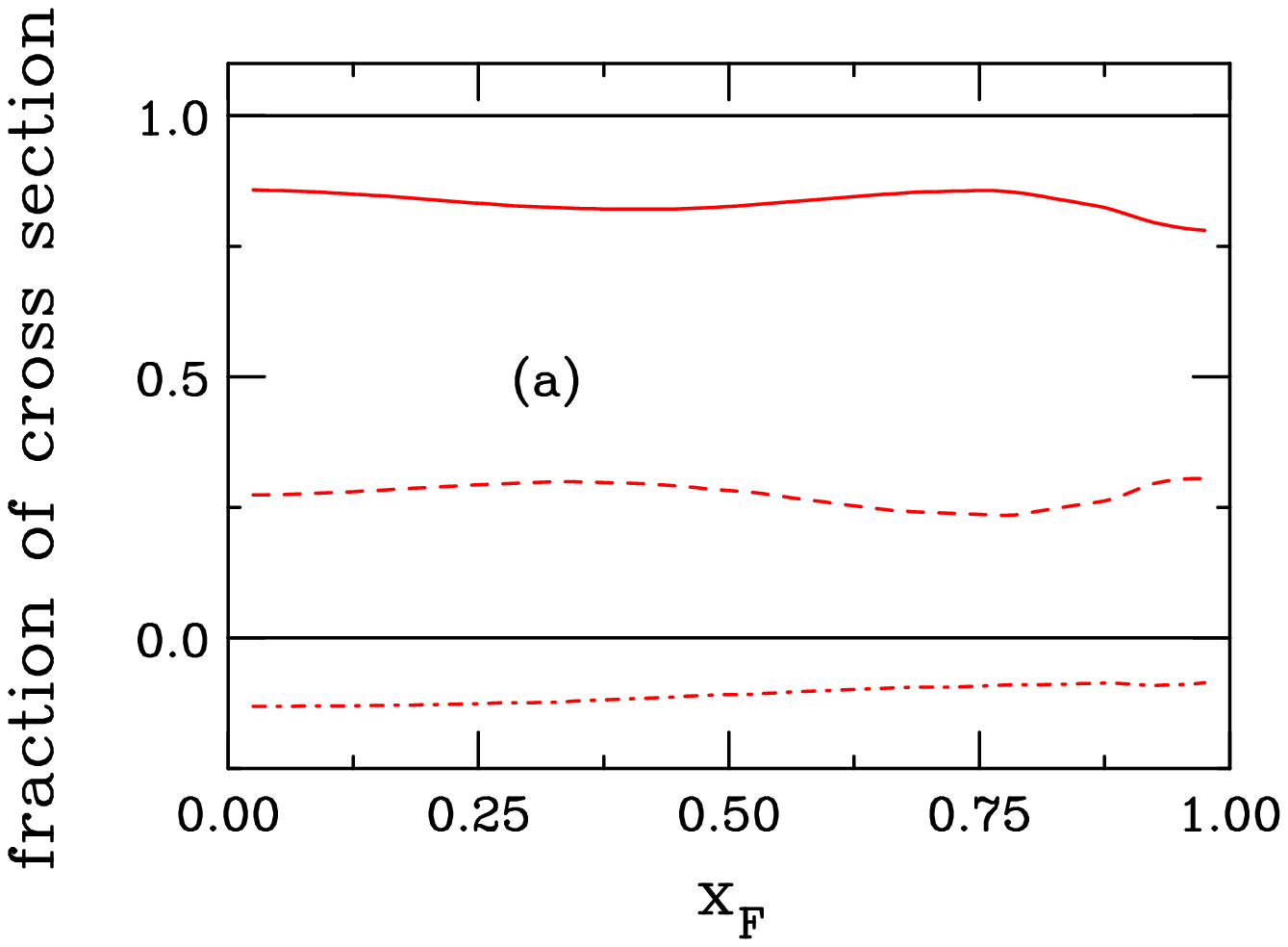}    
    \includegraphics[width=0.495\textwidth]{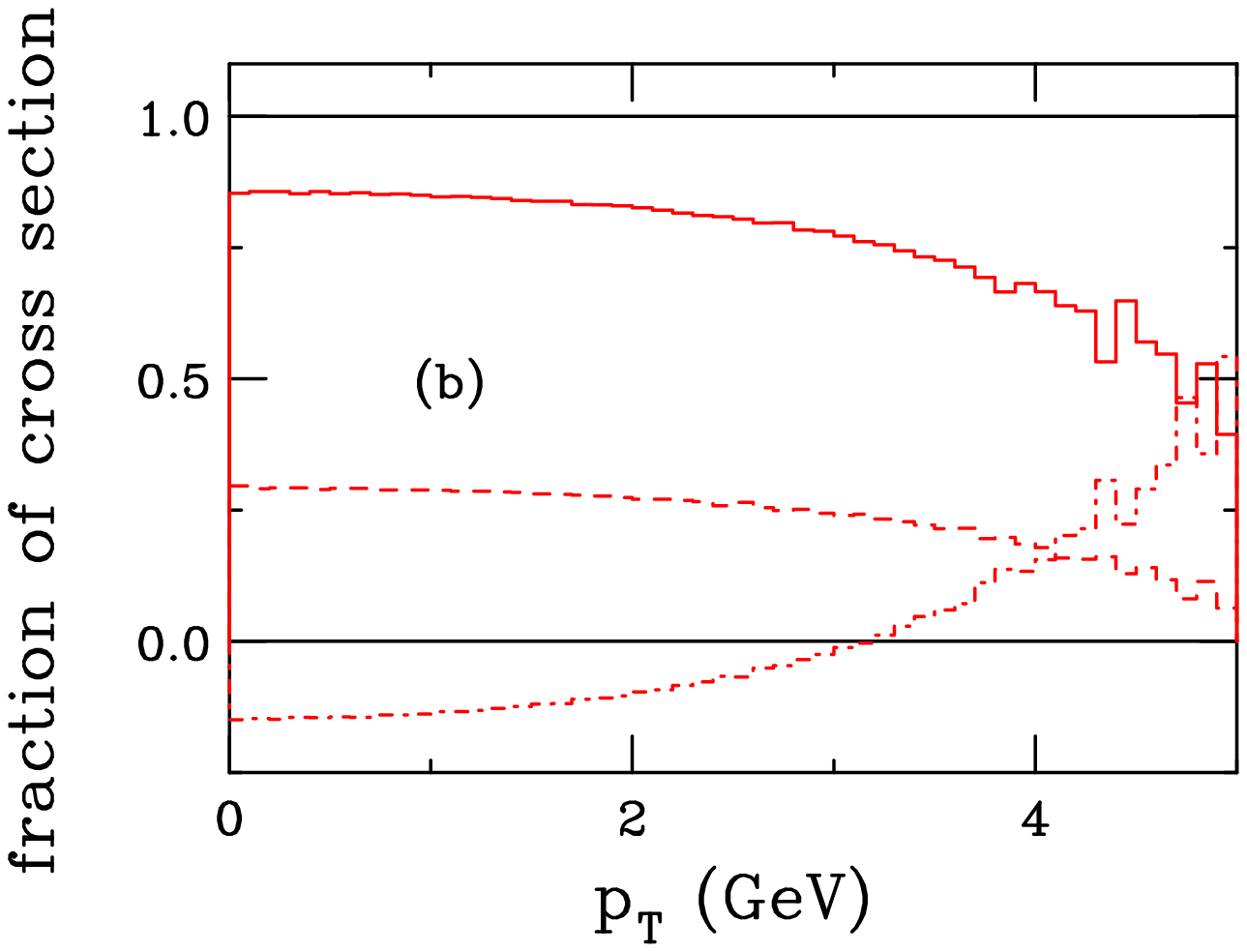}        
  \end{center}
  \caption[]{(Color online) The contributions to the $J/\psi$ production
    cross sections in the CEM in $p+p$ collisions at $\sqrt{s} = 15.4$~GeV
    as a function of $x_F$ (a) and $p_T$ (b) are shown for the $gg$ (solid),
    $q\overline q$ (dashed) and $qg$ (dot-dashed) initial states.}
\label{pp_contribs}
\end{figure}

Note that in the acceptance of the SeaQuest experiment, $0.4 < x_F < 0.95$
and $p_T < 2.3$~GeV, $\sim 22$\% of the total cross section is captured.  In
particular, the far forward acceptance in $x_F$ is in a region where one might
expect $q \overline q$ or $qg/\overline qg$ contributions to
heavy flavor production to be non-negligible compared to $gg$ production
of $J/\psi$.  Figure~\ref{pp_contribs} shows the ratio of the partial cross
sections for each partonic initial state, the squared amplitude weighted by the
parton densities, relative to the total cross section.
Indeed at $\sqrt{s_{NN}} = 15.4$~GeV, if $J/\psi$ production is
broken down into its $gg$, $q \overline q$ and $qg/\overline q g$ contributions,
the $gg$ contribution is $\approx 80$\% over the $x_F$ range, integrated over
all $p_T$, while the
$q\overline q$ and $qg$ contributions are $\sim 25$\% and
$\sim -10$\% respectively, see Fig.~\ref{pp_contribs}(a).
The contributions from the production channels
change more as a function of $p_T$, integrated over forward $x_F$, as shown
in Fig.~\ref{pp_contribs}(b), As $p_T$
increases the $gg$ and $qg$ contributions become comparable at $p_T \sim 5$~GeV
while the $q \overline q$ contribution decreases.
See Refs.~\cite{NDE1,MNR} for discussions of the relative contributions of the
production channels.
Also, note that even though the $qg$ contribution is negative over much of the
kinematic range, the total cross section is positive.
These results show that in the
specific kinematics and the low center of mass energy of the SeaQuest
experiment, the $gg$ contribution, while it is larger than the other
contributions, is
not fully dominant.  At higher energies, this component is larger than 95\%
due to both the larger $gg$ partonic cross section at high energies and the
large $gg$ luminosity at high energies and small $x$.

These
distributions are not significantly modified in $p+{\rm d}$ relative to $p+p$
collisions.  The $p+{\rm d}$
cross section is about 1-2\% larger per nucleon than the
$p+p$ cross section.  This difference can be attributed to the modified valence
quark contributions in proton relative to deuterium targets and is
non-negligible at the SeaQuest energy because the $q \overline q$ contribution
is around 25\% of the cross section in the SeaQuest acceptance.

\subsection{Nuclear Effects on the Parton Densities}
\label{shad}

The parton distribution functions in nuclei are modified from those of a free
proton.  A characteristic dependence of this modification has been determined
from deep inelastic scattering experiments of leptons (electrons, muons, or
neutrinos) on nuclei.  These leptonic measurements probe only charged partons,
leaving the gluon modifications to be inferred through the evolution of the
parton densities with scale and momentum sum rules.  The characteristic
dependence is usually presented as the ratio of the structure function in
a heavier nucleus, $F_{2 A}$ relative to the structure function in the
deuteron, $F_{2 {\rm d}}$.
The deuteron is chosen because it has an equal number of
up and down valence quarks between the constitutent neutron and proton.  At high
momentum fractions, $x > 0.3$, carried by the interacting parton, there is a
depletion in the nucleus relative to a nucleon, known as the EMC Effect.  At
smaller values of $x$, $x < 0.03$, there is also a depletion, known as
shadowing.  In the intermediate range of $x$, bridging the two regions where
the ratio $R = F_{2 A}/F_{2 {\rm d}} < 1$,
there is an enhancement referred to as
antishadowing.  

A number of global analyses have been made over time by several groups to
describe the modification as a function of $x$ and factorization scale $\mu_F$,
assuming collinear factorization and starting from a minimum scale, $\mu_F^0$.
These analyses have evolved with time, similar to global analyses of the proton
parton distribution functions, as more data become available.

Nuclear PDF effects are
generally implemented by a parameterization of the modification as
a function of $x$, $\mu_F$ and $A$ so that the $k_T$-independent
proton parton distribution
function in Eq.~(\ref{sigCEM}) is replaced by the nuclear parton distribution
function
\be
f_j^A(x_2,\mu_F^2) = R_j(x_2,\mu_F^2,A) f_j^p(x_2,\mu_F^2)
\ee
Note that the separate modifications for valence and sea quarks in the case of
up and down quarks requires the modification to be applied separately to the
valence and sea quark proton distributions.  The isospin also needs to be taken
into account for the light quark distributions as well to distinguish between
proton and neutron contributions.

Here the
EPPS16 \cite{EPPS16} nPDF parameterization at next-to-leading order in
the strong coupling constant $\alpha_s$ is employed in the calculations.
It is available for $A = 12$ (carbon), 56 (iron) and 184 (tungsten) but assumes
no modification for the deuteron target.

The EPPS16 analysis used the LHC data from the 2012 $p+ {\rm Pb}$ run at
$\sqrt{s_{NN}}= 5.02$~TeV.  Data on dijet production, as well as for
$W^\pm$ and $Z$ gauge boson data, are
still rather sparse but probe much higher scales than previously available from
fixed-target nuclear deep-inelastic scattering experiments.
This analysis also included
the CHORUS data \cite{50} with neutrino and antineutrino beams.  All these
additional data sets gave access to a wide range of scales,
$4 \leq \mu_F^2 \leq 2 \times 10^4$~GeV$^2$, at relatively
large $x$.  They were also sensitive to differences in the valence and sea
quark distributions, allowing separation between $u_V$ and $d_V$ and
$\overline u$ and $\overline d$ respectively.

EPPS16 has 20 fit parameters, giving
41 total sets with one central set and 40 error sets.  The error sets are
determined by varying each parameter individually within one standard deviation
of its best fit value.
The uncertainties on the individual quark ratios are calculated by summing the
excursions of each of the error sets from the central value in quadrature. 
The sets cover the range $1.3 < Q < 10000$~GeV and $10^{-7} < x < 1$.

The uncertainties on the $J/\psi$ distributions due to EPPS16 are obtained by
calculating the $J/\psi$ cross sections with the central set and the 40 error
sets and summing the differences in quadrature.  The resulting $J/\psi$
uncertainty band deviates from the central cross section on the order of 20\%,
significantly less than the mass and scale uncertainties shown in
Fig.~\ref{cem_pp}.

\begin{figure}
  \begin{center}
    \includegraphics[width=0.495\textwidth]{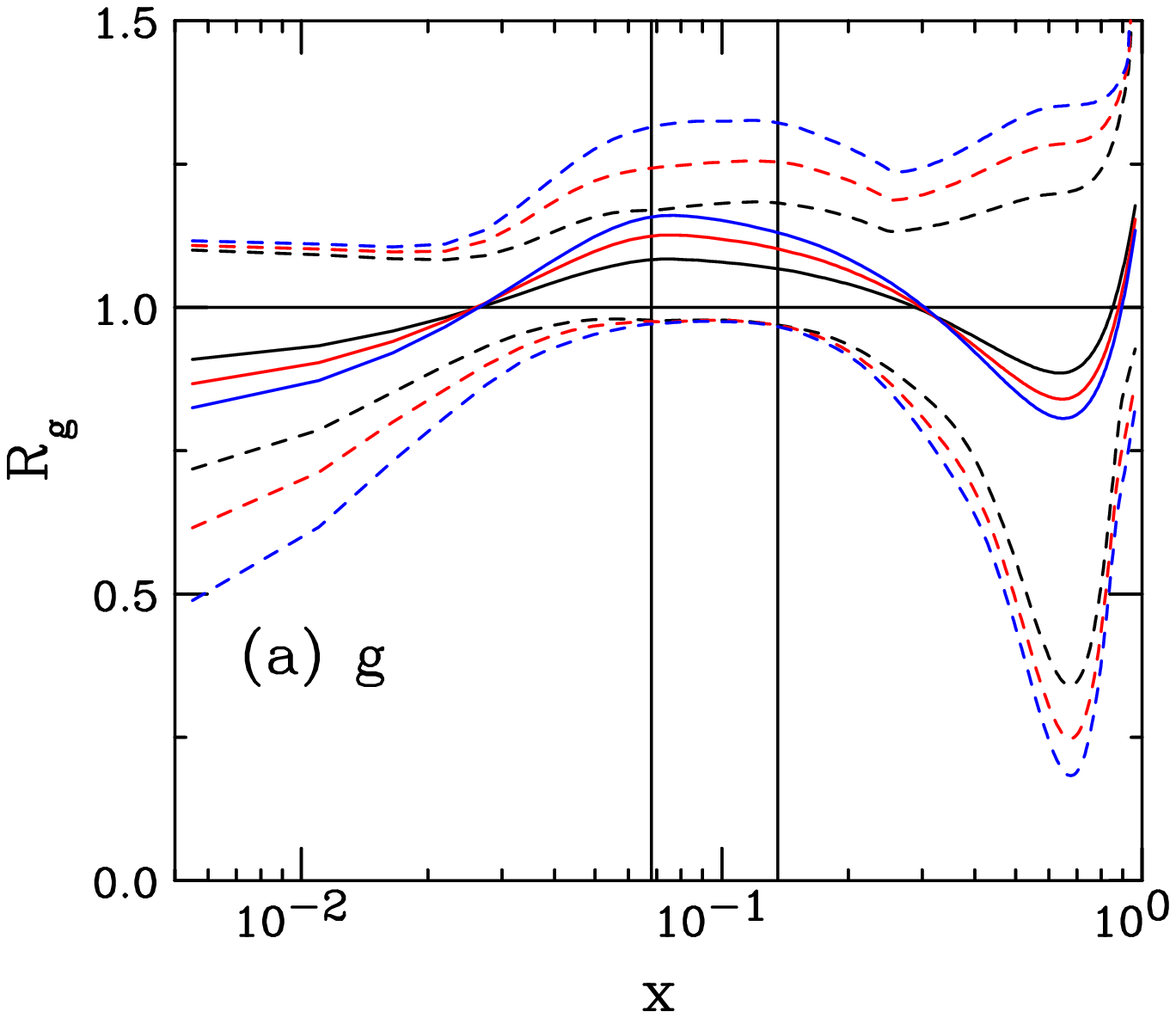}    
    \includegraphics[width=0.495\textwidth]{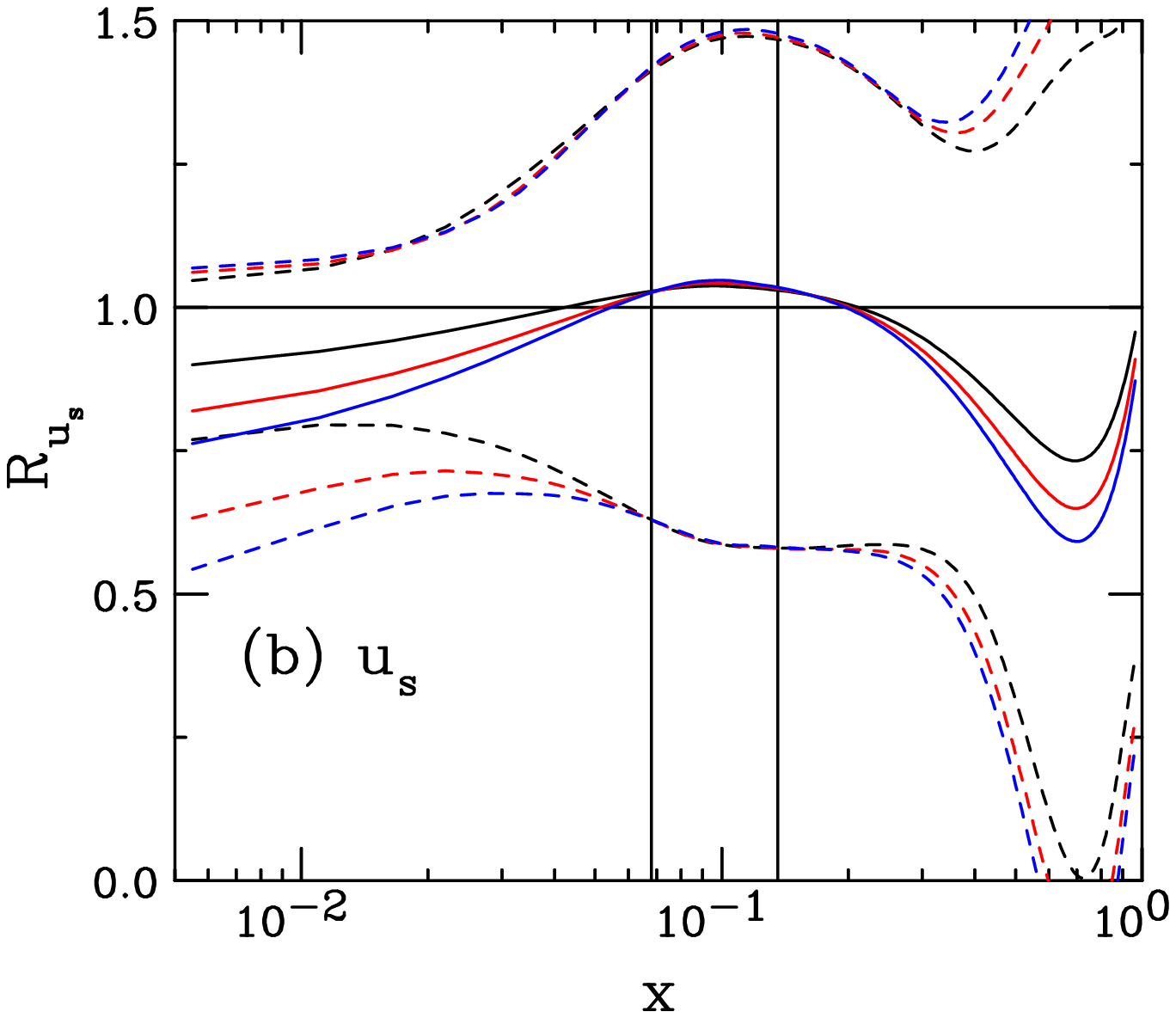}        
    \includegraphics[width=0.495\textwidth]{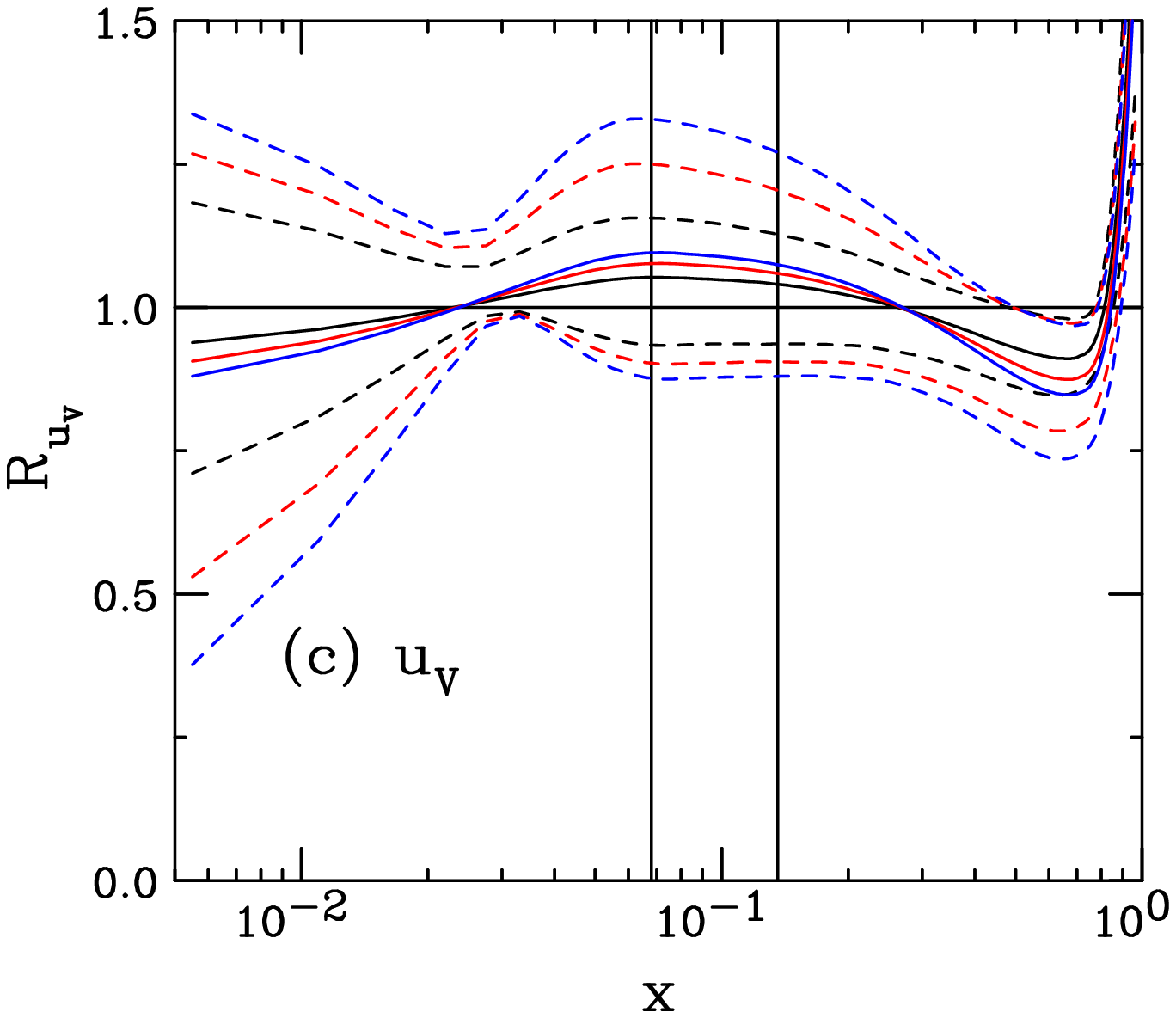}        
  \end{center}
  \caption[]{(Color online) The EPPS16 ratios, with uncertainties,
    are shown at the scale of the $J/\psi$ mass for gluons (a), the up sea quark
    distribution (b) and the valence up quark distribution (c) as a function of
    momentum fraction $x$. The central set is denoted by the solid curves
    while the dashed curves give the upper and lower limits of the uncertainty
    bands.  The results are given for $A = 12$ (black), 56 (red) and 184
    (blue).  The vertical lines indicate the $x$ range of the SeaQuest
    measurement, $0.068 < x < 0.136$.
  }
\label{shad_ratios}
\end{figure}

The EPPS16 ratios for gluons, valence up quarks and $\overline u$ quarks are
shown at the $J/\psi$ mass scale in Fig.~\ref{shad_ratios}.
The central sets, along with the uncertainty
bands, are shown for all three targets for gluons (a), anti-up quarks (b) and
valuence up quarks (c).  The range $x > 0.007$ is emphasized since SeaQuest
does not probe the small $x$ region.  Indeed, the average $x$ range probed
by SeaQuest is rather narrow and in the range where antishadowing is
dominant.  To set the limits in $x$ shown on Fig.~\ref{shad_ratios}, the
momentum fraction $x_2$ in the nucleus is calculated for $x_F = 0.95$, $p_T = 0$
and $x_F = 0.4$, $p_T = 2.3$~GeV in $2 \rightarrow 2$ kinematics where
$x_2 = 0.5(-x_F + (x_F^2 + 4m_T^2/s)^{1/2})$ and
$m_T = (m_{J/\psi}^2 + p_T^2)^{1/2}$,
assuming that any light final-state parton associated with $J/\psi$ production
in the $2 \rightarrow 3$ next-to-leading order kinematics has a small effect on
the resulting $x_2$.
These combinations of $x_F$ and $p_T$ give the minimum and maximum values of
$x_2$, 0.068 and 0.136 respectively.

The ratios are all shown for the $J/\psi$ mass,
$\mu_F = m_{J/\psi}$.  At this value of the factorization
scale, within a factor of three of
the minimum scale employed in the EPPS16 fits, $\mu_F = 1.3$~GeV, the
uncertainty band is quite large.

The fixed target neutrino deep-inelastic scattering data and the gauge boson
data from the LHC reveal differences between the valence and sea quark ratios.
The nuclear modification is
weaker for $d_V$ than  for $u_V$ while there is no antishadowing
for $\overline d$.  On the
other hand, the $\overline u$ ratio shows antishadowing on a level similar to
the valence quarks, albeit over a narrower range in $x$.  Only the $\overline u$
and $u_V$ modifications are shown in Fig.~\ref{shad_ratios}.  Examples of the
sea quark and valence quark ratios are
shown because SeaQuest is in a kinematic region where $J/\psi$ production is
not dominated by production via the $gg$ channel, see Fig.~\ref{pp_contribs}.

Finally, it is worth noting that these modifications assume minimum bias
collisions, averaging over the volume of the nucleus.  Measurements of $J/\psi$
production in $p+{\rm Pb}$ collisions at much lower
$x$ such as of the ALICE Collaboration, have been shown as a function of
collision centrality, as determined by event multiplicity \cite{ALICE_QpPb}.
These results exhibit a clear dependence on where the probe impacts the nucleus,
for example, in peripheral collisions, closer to the `edge' of the nucleus
where there might be only one nucleon in the path of the proton or more
central collisions where the proton may encounter multiple
nucleons \cite{Kitagaki:1988wc,Emelyanov:1999pkc}.  It would be interesting to
see whether similar modifications could be observed in high-statistics
fixed-target measurements such as at SeaQuest.

\subsection{Nuclear Absorption in $pA$ Interactions}
\label{absorption}

In $p+A$ collisions, the proto-$J/\psi$ produced in the initial parton
interactions may further interact with nucleons and be dissociated or
absorbed before it can escape the target, referred to as nuclear absorption.
The effect of nuclear absorption alone on the $J/\psi$ 
production cross section in
$p+A$ collisions may be expressed as \cite{rvrev}
\begin{eqnarray}
  \sigma_{pA} = \sigma_{pN} S_A^{\rm abs} & = & \sigma_{pN} \int d^2b \,
  \int_{-\infty}^{\infty}\, dz \, \rho_A (b,z) S^{\rm abs}(b) \\
& = & \sigma_{pN} \int d^2b \,  \int_{-\infty}^{\infty}\, dz \, \rho_A (b,z)
 \exp \left\{
-\int_z^{\infty} dz^{\prime} \rho_A (b,z^{\prime}) \sigma_{\rm abs}(z^\prime
-z)\right\} \, \, , 
\label{sigfull}
\end{eqnarray} where $b$ is the impact parameter, $z$ is the $c \overline c$
production point, $S^{\rm abs}(b)$ is the nuclear absorption survival 
probability, and $\sigma_{\rm abs}(z^\prime -z)$ 
is the nucleon absorption cross section.  Evem though
the absorption cross section
used in this work is assumed to be constant, it is written as a function of the
path length through the nucleus in Eq.~(\ref{sigfull}) because other functional
forms have previously been assumed, see {\it e.g.}\ Ref.~\cite{RV_HeraB}.
Expanding the exponent in Eq.~(\ref{sigfull}), integrating, 
and reexponentiating the results assuming $A$ is large
leads to Eq.~(\ref{alfint}) with $S_A^{\rm abs} = A^\alpha$ and
$\alpha = 1 - 9\sigma_{\rm abs}/(16 \pi r_0^2)$ \cite{rvrev}.

There have been many studies of the absorption mechanism of the $J/\psi$ in
nuclei and the cross section obtained in various model studies depends on
whether the $J/\psi$ is assumed to be produced inside or outside the target and
whether it is produced as a color singlet or color octet.  For comparison
of results assuming different absorption scenarios, see for example
Refs.~\cite{RV_HeraB,RV_e866,LWV,Darren} and references therein.

The extracted absorption cross section also depends on which cold nuclear matter
effects were taken into account.  If one assumes only absorption, the cross
section could be underestimated at low energies because antishadowing effects
have been neglected.  At energies where antishadowing dominates, a larger
effective absorption cross section is needed to reach the level of measured
suppression in $p+A$ relative to $p+p$ collisions, as discussed in
Ref.~\cite{LWV}.  In that reference, the effective absorption cross section
was extracted for each measured $x_F$ or rapidity point for several nPDF
parameterizations available at the time.  While the effective absorption
showed some dependence on the chosen nPDF parameterization, the authors
were able to discern some trends.  At midrapidity or $x_F \sim 0$, absorption
was generally largest and was seen to decrease forward of this central region,
at least up to a point.  Experiments such as Fermilab's E866 experiment
\cite{e866} at $\sqrt{s_{NN}} = 38.8$~GeV or the NA3 collaboration
\cite{NA3} at CERN with $\sqrt{s_{NN}} = 19.4$~GeV showed a strong increase in
effective absorption for more forward $x_F$, $x_F > 0.3$ \cite{LWV}.  It was
also noted, including by the E866 Collaboration \cite{e866}, that although
high $x_F$ also corresponds to low $x_2$ and the shadowing region, results at
different energies did not scale with $x_2$, demonstrating that shadowing
could not be the dominant effect in that region.  Other effects such as energy
loss in cold matter were speculated
to lead to the observed behavior.

Based on the midrapidity results, an energy-dependent effective absorption
cross section was obtained employing the available data \cite{LWV}.
Extrapolating the findings
back to 120~GeV gives $7 \leq \sigma_{\rm abs} \leq 9$~mb.
A value of 9~mb is used here and is compared to results with no nucleon
absorption, $\sigma_{\rm abs} = 0$ and $\alpha = 1$.

Even though $J/\psi$ absorption may be inextricably entangled with other cold
nuclear matter effects in hadroproduction,
it may be possible to make an independent measurement of the $J/\psi$-nucleon
total cross section, effectively the absorption cross section, from measurement
of the $A$ dependence in photoproduction.  Such an experiment was
performed at SLAC at 20~GeV \cite{SLACphoto}, giving a value of
$3.5 \pm 0.8 \pm 0.5$~mb, lower than estimates from hadroproduction at
higher energies \cite{LWV}.  A new experiment of the $A$ dependence at similar
energies would be useful \cite{StanChudakov}, especially if the coherent and
incoherent contributions could be separated.  Measurements of $J/\psi$
production in ultraperipheral collisions, while not shedding new light on 
absorption by nucleons, can provide additional constraints on the nuclear gluon
parton densities.  Measurements reported by ALICE \cite{32,33} and
CMS \cite{CMS_UPC} have already put some constraints on the gluon density in
the nucleus at low $x$.  

Absorption by comoving particles, whether characterized
as hadrons or partons, has been included in some calculations of cold nuclear
matter effects \cite{Elena}.  This type of suppression
has not been included separately
here because it has been shown to have the same nuclear dependence in minimum
bias collisions \cite{SGRV1990}.

\subsection{$k_T$ Broadening}
\label{kTkick}

The effect and magnitude of intrinsic $k_T$ broadening on the $J/\psi$ $p_T$
distribution in $p+p$ collisions was discussed in Sec.~\ref{CEM}.
Here further broadening due to the
presence of a nuclear target is considered.
One might expect that a higher intrinsic $k_T$
is required in a nuclear medium relative to that in $p+p$ due to multiple
scattering in the nucleus, known as the Cronin effect \cite{Cronin}.  The
effect is implemented by replacing $g_p(k_T)$ in Eq.~(\ref{FSgauss}) by
$g_A(k_T)$ where $\langle k_T \rangle_p$ is replaced by $\langle k_T \rangle_A$
in Eq.~(\ref{intkt}).

The total broadening in a nucleus relative to a nucleon can be expressed as
\begin{eqnarray}
\langle k_T^2 \rangle_A = \langle k_T^2 \rangle_p +\delta k_T^2 \, \, .
\end{eqnarray}
The same expression for $\delta k_T^2$ used in Ref.~\cite{HPC_pA}, based on
Ref.~\cite{XNW_PRL}, is employed here,
\begin{eqnarray}
  \delta k_T^2 = (\langle \nu \rangle - 1) \Delta^2 (\mu) \, \, .
  \label{delkt2}  
\end{eqnarray}
The strength of the broadening, $\Delta^2 (\mu)$, depends on the interaction
scale \cite{XNW_PRL},
\begin{eqnarray}
  \Delta^2 (\mu) = 0.225 \frac{\ln^2 (\mu/{\rm GeV})}{1  + \ln(\mu/{\rm GeV})}
  {\rm GeV}^2 \, \, .
\label{delta2}
\end{eqnarray}
The scale at which the broadening is evaluated is taken to be $\mu = 2m_c$
\cite{HPC_pA}.  The scale dependence suggests that one might expect a
larger $k_T$ kick in the nucleus for bottom quarks than charm quarks.

The size of the effect depends on the number of scatterings the incident proton
could undergo while passing through a nucleus.  This is given by
$\langle \nu \rangle -1$, the number of collisions, less the first collision.
The number of scatterings
strongly depends on the impact parameter of the proton with respect to the
nucleus.  However, in minimum-bias collisions, the impact parameter is
averaged over all possible paths.  The average number of scatterings is
\begin{eqnarray}
  \langle \nu \rangle
  = \sigma_{pp}^{\rm in} \frac{\int d^2b T_A^2(b)}{\int d^2b T_A(b)}
  = \frac{3}{2} \rho_0 R_A \sigma_{pp}^{\rm in}
  \label{avenu}
\end{eqnarray}
where $T_A(b)$ is the nuclear profile function,
$T_A(b) = \int_{-\infty}^{\infty} dz \rho_A(b,z)$
and $\rho_A$ is the nuclear density distribution.
An average nuclear density, $\rho_0 = 0.16$/fm$^3$, is assumed for a spherical
nucleus of radius $R_A = 1.2 A^{1/3}$; and $\sigma_{pp}^{\rm in}$ is
the inelastic $p+p$ cross section, $\sim 32$~mb at fixed-target energies.  

With the given values of $\rho_0$ and $\sigma_{pp}^{\rm in}$ in Eq.~(\ref{avenu})
and $\Delta^2(\mu = 2m_c) = 0.101$~GeV$^2$ from Eq.~(\ref{delta2}),,
\begin{eqnarray}
  \delta k_T^2 \approx (0.92 A^{1/3} - 1)\times 0.101 \, {\rm GeV}^2 \, \, .
  \end{eqnarray}
For carbon, iron and tungsten targets, $\delta k_T^2 = 0.1$, 0.25, and
0.39~GeV$^2$ respectively.  This gives an average broadening in the nucleus of
$\langle k_T^2 \rangle_A = 1.07$, 1.22, and 1.36~GeV$^2$ for the SeaQuest
targets.  The change in $\langle k_T^2 \rangle$ between carbon and tungsten
targets is similar to that observed for $J/\psi$ production in other
fixed-target experiments \cite{herab2}.  That work also observed that this
relative broadening was independent of center-of-mass energy.

The effect of $k_T$ broadening in nuclei, relative to the proton, is to reduce
the ratio $p+A/p+p$ or $p+A/p+{\rm d}$ at low $p_T$ and enhance it at high
$p_T$.  Because $x_F$ depends on $m_T$, one can expect a small effect as a
function of $x_F$ while a more significant one should be seen as a function
of $p_T$.

\section{Intrinsic Charm}
\label{ICcomp}

The wave function of a proton in QCD can be represented as a
superposition of Fock state fluctuations, {\it e.g.}\ $\vert uudg
\rangle$, $\vert uud q \overline q \rangle$, $\vert uud Q \overline Q \rangle$,
\ldots of the $\vert uud \rangle$ state.
When the projectile scatters in the target, the
coherence of the Fock components is broken and the fluctuations can
hadronize \cite{intc1,intc2,BHMT}.  These
intrinsic $Q
\overline Q$ Fock states are dominated by configurations with
equal rapidity constituents. Thus the intrinsic heavy quarks carry a large
fraction of the projectile momentum \cite{intc1,intc2}.
 
The frame-independent probability distribution of a $5$-particle
$c \overline c$ Fock state in the proton is 
\be
dP_{{\rm ic}\, 5} = P_{{\rm ic}\,5}^0
N_5 \int dx_1 \cdots dx_5 \int dk_{x\, 1} \cdots dk_{x \, 5}
\int dk_{y\, 1} \cdots dk_{y \, 5} 
\frac{\delta(1-\sum_{i=1}^5 x_i)\delta(\sum_{i=1}^5 k_{x \, i}) \delta(\sum_{i=1}^5 k_{y \, i})}{(m_p^2 - \sum_{i=1}^5 (\widehat{m}_i^2/x_i) )^2} \, \, ,
\label{icdenom}
\ee
where $i = 1$, 2, 3 are the interchangeable light quarks ($u$, $u$, $d$)
and $i = 4$ and 5 are the $c$ and $\overline c$ quarks respectively.
Here $N_5$ normalizes the
$|uud c \overline c \rangle$ probability to unity and $P_{{\rm ic}\, 5}^0$
scales the unit-normalized
probability to the assumed intrinsic charm content of the proton.  The delta
functions conserve longitudinal ($z$) and transverse ($x$ and $y$) momentum.
The denominator of Eq.~\ref{icdenom} is
minimized when the heaviest constituents carry the largest fraction of the
longitudinal momentum, $\langle x_Q \rangle > \langle x_q \rangle$.

Note that only this 5-particle Fock state of the proton is
considered here since it gives the most forward $x_F$ production of $J/\psi$
from intrinsic charm.  One can also consider higher Fock components such as
$|uud c\overline c q \overline q \rangle$ but these will reduce the average
momentum fraction of the $J/\psi$ and also enter at lower probabilities, see
e.g.\ Refs.~\cite{tomg} for examples of charm hadron distributions from higher
Fock states.

Heretofore, only the $x_F$ dependence of the $J/\psi$ from the 5-particle
Fock state has been considered.  In this case average values for the transverse
masses of the constituent quarks,
$\widehat{m}_i^2 = m_i^2 + k_{T \, i}^2$, can be used.  Previously,
$\widehat{m}_q = 0.45$ GeV and $\widehat{m}_c = 1.8$ GeV were chosen
\cite{VBH1}.  In this case, the $J/\psi$ $x_F$ distribution can be calculated
assuming simple coalescence of the $c$ and $\overline c$ in a single state,
represented in 
Eq.~(\ref{icdenom}) by the addition of a delta function,
$\delta(x_F - x_c - x_{\overline c}$, for the longitudinal, $z$, direction.  We
can related the summed $x_c$ and $x_{\overline c}$ momentum fractions to the $x_F$
of the $J/\psi$ assuming that the it is brought on-shell by a soft scattering
with the target.

In this work, the $p_T$ distribution of the $J/\psi$ from such a Fock state is
calculated explicitly for the first time.  Thus the $k_x$ and $k_y$ integrals
are also represented in Eq.~(\ref{icdenom}).  To simplify the calculations, it
is assumed that the transverse momentum of the $J/\psi$, $p_T$, is in the $x$
direction only.  This introduces two delta functions to Eq.~(\ref{icdenom}),
$\delta(p_T - k_{x \, c} - k_{x \, \overline c})$ and
$\delta(k_{y \, c} + k_{y\, \overline c})$.  

Employing these delta functions, the longitudinal momentum fraction of the  
$\overline c$ can be integrated, as well as the transverse momentum of the 
$\overline c$ quark in the proton, leaving the light quark $x$ and $\vec k$
as well as that of the $c$ quark to be integrated numerically.  

\begin{figure}
  \begin{center}
    \includegraphics[width=0.495\textwidth]{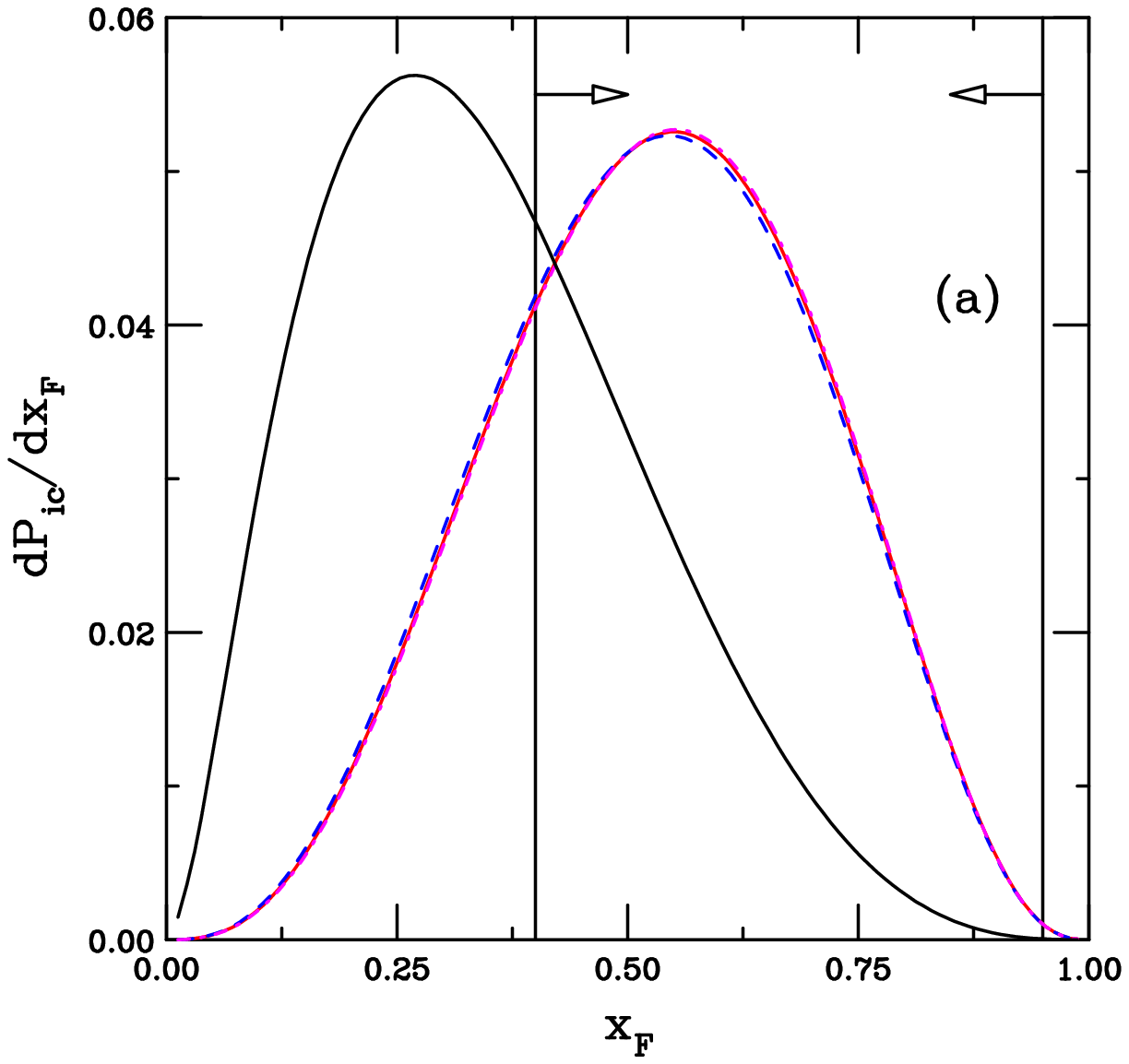}
    \includegraphics[width=0.495\textwidth]{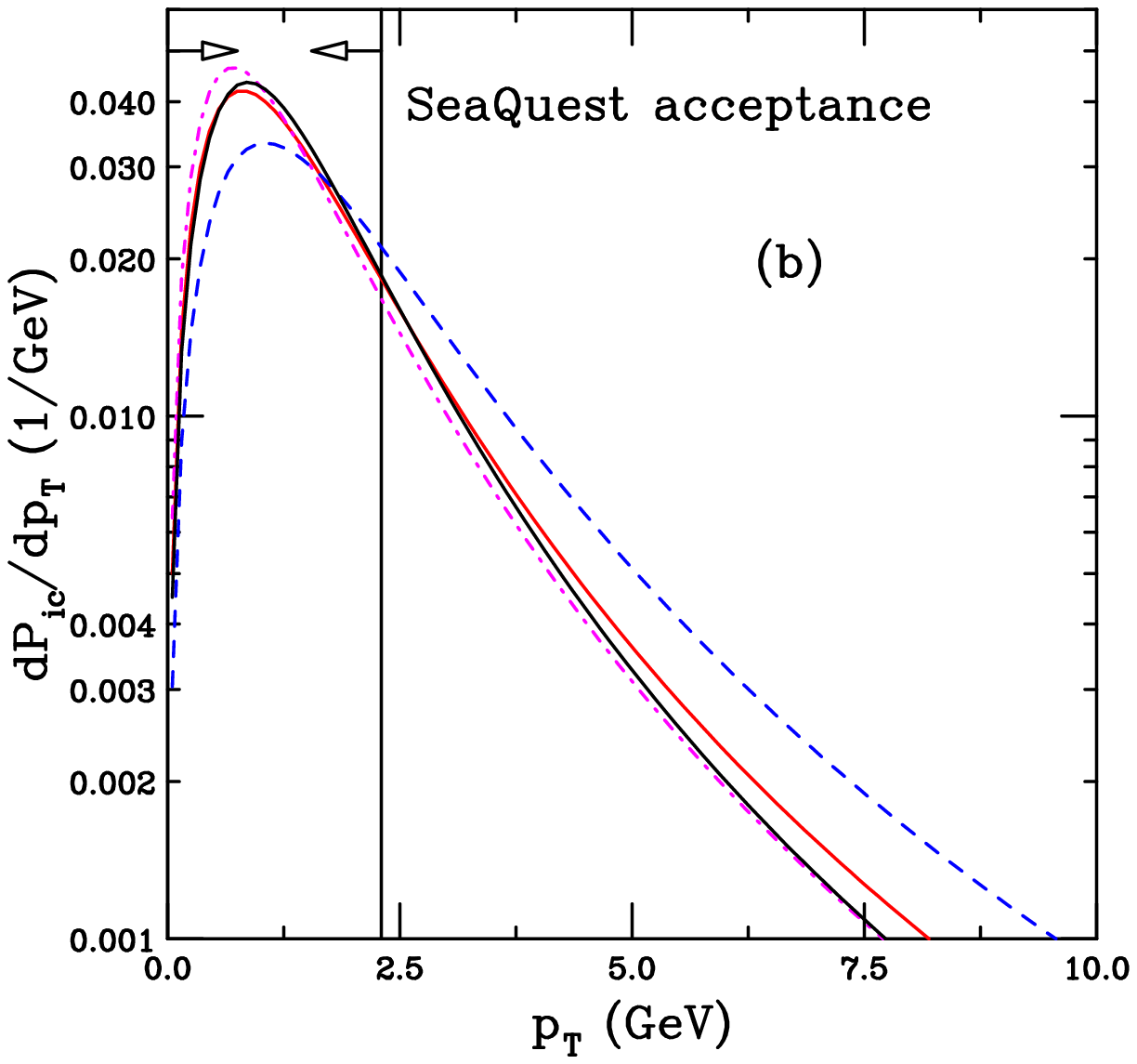}
  \end{center}
  \caption[]{(Color online) The probability distributions for $J/\psi$
    production from a five-particle proton Fock state as a function of $x_F$
    (a) and $p_T$ (b).  The results are shown for different values of the $k_T$
    range for the light and charm quarks.  The red curve employs the default
    values, $k_q^{\rm max} = 0.2$~GeV and $k_c^{\rm max} = 1.0$~GeV while the blue
    dashed curve increases $k_q^{\rm max}$ and $k_c^{\rm max}$
    by a factor of two and the dot-dashed
    magenta curve employs half the values of $k_q^{\rm max}$ and $k_c^{\rm max}$.
    The solid black
    curve shows the $x$ and $p_T$ distributions for a single charm quark from
    the state.}
\label{ic_dists}
\end{figure}

The $J/\psi$ $x_F$ and $p_T$ probability distributions
from intrinsic charm are shown in Fig.~\ref{ic_dists}.  The distributions are
normalized to unity.  How this normalized probability distribution is
converted into a cross section for $J/\psi$ production by intrinsic charm is
explained shortly.
Results are shown for different values of
$k_T$ for the light and charm quarks in the proton.  

As shown in Fig.~\ref{ic_dists}(a), the $x_F$ 
dependence is effectively independent of the chosen $k_T$ limits.  The
average $x_F$ of the distribution is 0.53, almost exactly the same as
previous calculations assuming constant values of $\hat{m}_q = 0.45$~GeV
and $\hat{m}_c^2 = 1.8$~GeV \cite{VBH1}.  The charm quark $x_c$ distribution
obtained by integrating over the momenta of the other four partons in the Fock
state, assuming the quark is liberated from the projectile proton by a soft
interaction.  It has an average value of $x_c = 0.34$,
slightly larger than half that of the $J/\psi$.

The $p_T$ distributions, on the other hand, show a significant dependence on
the range of $k_T$ integration.  The chosen default values are
$k_q^{\rm max} = 0.2$~GeV and $k_c^{\rm max} = 1.0$~GeV, shown by the red curve
in Fig.~\ref{ic_dists}(b).  Note that these limits apply to both the $x$ and
$y$ directions of the transverse momentum.
The charm quark distribution, with these same
default values for the range of integration, is given by the black curve.

If the limits of the integration range are doubled to $k_q^{\rm max} = 0.4$~GeV
and $k_c^{\rm max} = 2.0$~GeV, the $p_T$ distribution is broadened, see the blue
dashed curve in Fig.~\ref{ic_dists}(b), while halving the integration range to
0.1~GeV and 0.5~GeV for the light and charm quarks respectively, results in
the dot-dashed magenta curve.  The average $J/\psi$ $p_T$ resulting from these
values of $k_i^{\rm max}$ are 2.06, 2.48, and 1.90~GeV for the default range,
doubling the range, and halving the range, respectively.  The average $p_T$ of
the charm quark in the default integration range is 2.00~GeV.

It is notable that the $J/\psi$ $p_T$ distribution from intrinsic charm is
considerably broader than that of the $J/\psi$ from the CEM shown in
Fig.~\ref{cem_pp}(b).  This can be easily understood, even when an average
$\langle k_T \rangle^2 \sim 1$~GeV$^2$ is employed with the CEM, when one
considers the two sources.

In perturbative QCD, the $p_T$ range depends on
the center of mass energy with a limit of $p_T \sim \sqrt{s_{NN}}/2$ at
$x_F = 0$ for
massless partons and less for massive quarks, where $p_T$ is replaced by $m_T$.
Leading order $2 \rightarrow 2$ scattering is also assumed for this estimate,
leading to $p_T \leq 7.2$~GeV in the massless case and $p_T \leq 6.5$~GeV
for the $J/\psi$ in the CEM.  In addition, since the initial partons taking
part in the interaction come from two different hadrons, one from the
projectile and the other from the target, at least one of them will carry a much
smaller fraction of the momentum that the intrinsic charm quarks in the proton
Fock state.

On the other hand, when the $J/\psi$ arises from an intrinsic charm state of
the proton, according to Eq.~(\ref{icdenom}), there is no energy limit on the
$p_T$ distribution other than that imposed by momentum conservation.  The
$J/\psi$ kinematics all come from the incident proton.  Only a soft interaction
with the target is sufficient to disrupt the Fock state and bring the $J/\psi$
on mass shell.

Even though intrinsic charm may dominate at higher $p_T$
for the SeaQuest energy and kinematics, at higher
energies, the $J/\psi$ cross section from perturbative QCD will be significantly
higher and the average $p_T$ of the CEM calculation grows with energy, along
with the intrinsic $k_T$ kick imparted to the $J/\psi$ at low $p_T$, see
Eq.~(\ref{eq:avekt}).
In particular, at colliders, the acceptance is generally around midrapidity
(low $x_F$) and the largest part of the intrinsic charm contribution is thus
outside the detector acceptance and detection of intrinsic charm is less
favorable.

The intrinsic charm production cross section from the
$|uudc \overline c \rangle$ component of the proton can be written as 
\be
\sigma_{\rm ic}(pp) = P_{{\rm ic}\, 5} \sigma_{p N}^{\rm in}
\frac{\mu^2}{4 \widehat{m}_c^2} \, \, .
\label{icsign}
\ee
The factor of $\mu^2/4 \widehat{m}_c^2$ arises from the soft
interaction which breaks the coherence of the Fock state where 
$\mu^2 = 0.1$~GeV$^2$ is assumed, see Ref.~\cite{VBH1}, and 
$\sigma_{pN}^{\rm in} = 30$~mb is 
appropriate for the SeaQuest energy.
Employing this cross section for charm production from the Fock state, the
$J/\psi$ contribution is assumed here to be calculated by scaling
Eq.~(\ref{icsign}) by the 
same factor, $F_C$, used to scale the CEM calculation to obtain the inclusive
$J/\psi$ cross section from the $c \overline c$ cross section in
Eq.~(\ref{sigCEM}),
\be
\sigma_{\rm ic}^{J/\psi}(pp) = F_C \sigma_{\rm ic}(pp) \, \, .
\label{icsigJpsi}
\ee
The nuclear dependence of the intrinsic charm contribution is assumed to be the
same as that extracted for the nuclear surface-like component of $J/\psi$
dependence by the NA3 Collaboration \cite{NA3} in Eq.~(\ref{twocom}), so that
\be 
\sigma_{\rm ic}^{J/\psi}(pA) = \sigma_{\rm ic}^{J/\psi}(pp) \, A^\beta \, \, 
\label{icsigJpsi_pA}
\ee
with $\beta = 0.71$ \cite{NA3} for a proton beam.

To represent the uncertainties on intrinsic charm, several values of the
intrinsic charm probability, $P_{{\rm ic}\, 5}^0$, are employed. 
The EMC charm structure function data is consistent with 
$P^0_{{\rm ic}\, 5} = 0.31$\%  for low energy virtual photons but
$P^0_{{\rm ic}\, 5}$
could be as large as 1\% for the highest virtual photon energies
\cite{EMC,hsv}.  For a lower limit, a probability of 0.1\% is used.
All three results are generally shown in Sec.~\ref{model_comp} unless otherwise
noted.

In this work, the formulation for intrinsic charm in the proton wavefunction
postulated by Brodsky and collaborators in Refs.~\cite{intc1,intc2} has been
adapted.  Other variants of the intrinsic charm distribution in the proton
exist, including meson-cloud models where the proton fluctuates into a
$\overline D(u \overline c) \Lambda_c (udc)$ state
\cite{Paiva:1996dd,Neubert:1993mb,Steffens:1999hx,Hobbs:2013bia} and a
sea-like distribution \cite{Pumplin:2007wg,Nadolsky:2008zw}.  
Intrinsic charm
has also been included in global analyses of the parton densities
\cite{Pumplin:2007wg,Nadolsky:2008zw,Dulat:2013hea,Jimenez-Delgado:2014zga,NNPDF_IC}.
The range of $P_{{\rm ic}\, 5}^0$
explored here is consistent with the results of the
global analyses.  For more details of these other works, see the review of
Ref.~\cite{IC_rev}.  New constraints from lattice gauge theory are also
available in Ref.~\cite{Sufian:2020coz}.  See also the recent review of
Ref.~\cite{Stan_review} for
more applications of intrinsic heavy quark states.

\section{Results}
\label{model_comp}

This section presents the results for the nuclear suppression factor,
\be
R_{pA} = \frac{2}{A} \frac{\sigma_{pA}}{\sigma_{p{\rm d}}} \, \, , \label{RpA_def}
\ee
where the cross section per nucleon in a nuclear target is compared to that of
the deuteron target.
The relative per nucleon cross sections are a more straightforward
way to present the nuclear dependence than using the exponent $\alpha$ as was
done previously.  Using the deuteron target reduces the potential, albeit 
small, isospin dependence of the ratios.

The calculations shown in this section are all in the $x_F$ and $p_T$
bins defined by SeaQuest \cite{Ayuso}.  The calculated points shown in the
figures are placed at the arithmetic center of the bin rather than a
cross-section weighted center.  Changing the location of the calculated points
within the bins could slightly modifiy the shape of the $x_F$ and $p_T$
dependence in this section but not enough to alter any conclusions.

Section~\ref{RpA:pQCD} shows the results with cold nuclear matter
effects on the perturbative QCD contribution alone while Sec.~\ref{RpA:pQCD+IC}
includes intrinsic charm, both without and with nuclear absorption on the
pQCD contribution.  Results are calculated both as a function of $x_F$ and
$p_T$ and are shown for all three nuclear targets: $A = 12$ (carbon);
$A = 56$ (iron); and $A=184$ (tungsten).

\subsection{$R_{pA}$ From Perturbative QCD Effects Alone}
\label{RpA:pQCD}

\begin{figure}
  \begin{center}
    \includegraphics[width=0.495\textwidth]{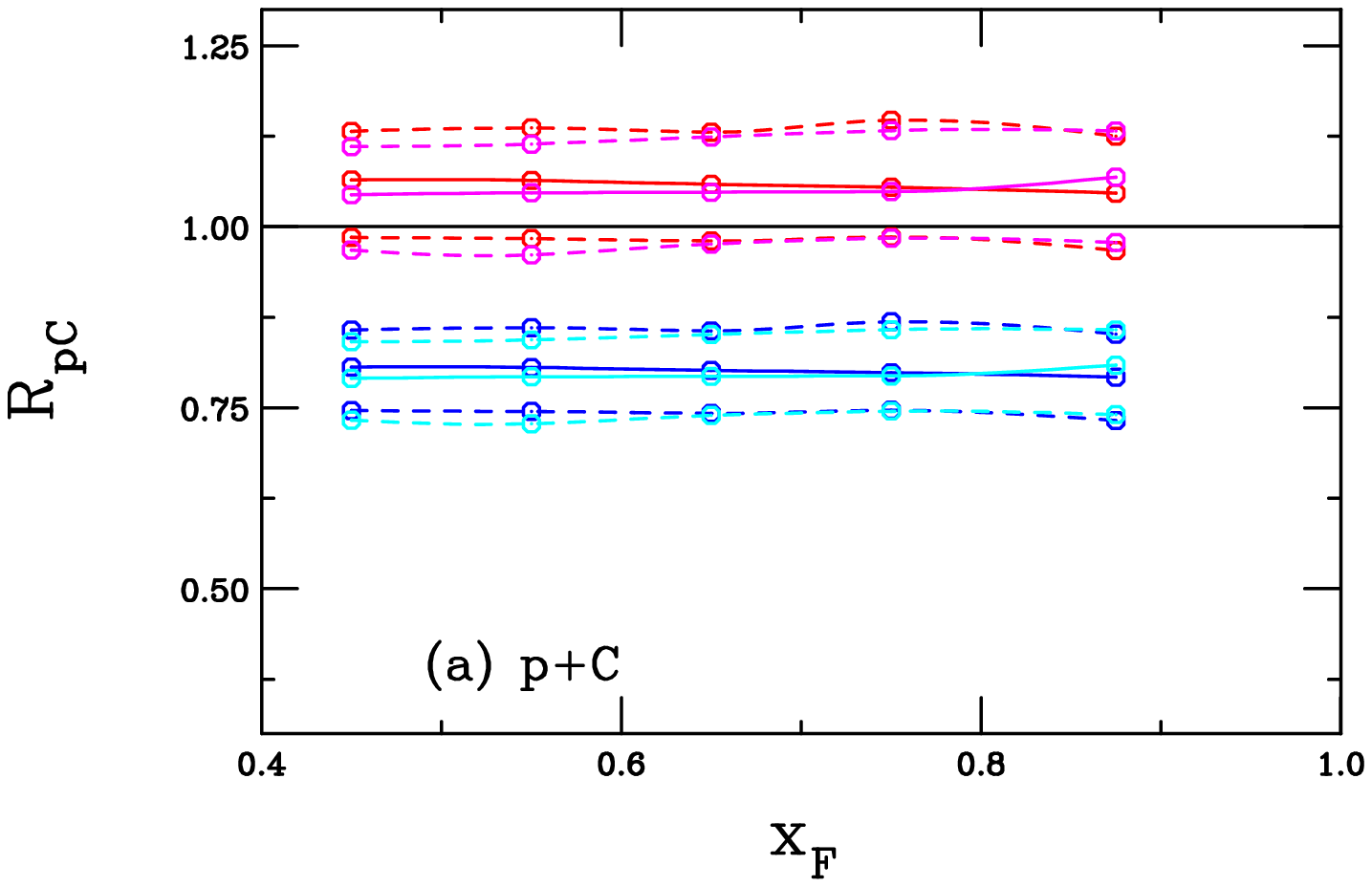}    
    \includegraphics[width=0.495\textwidth]{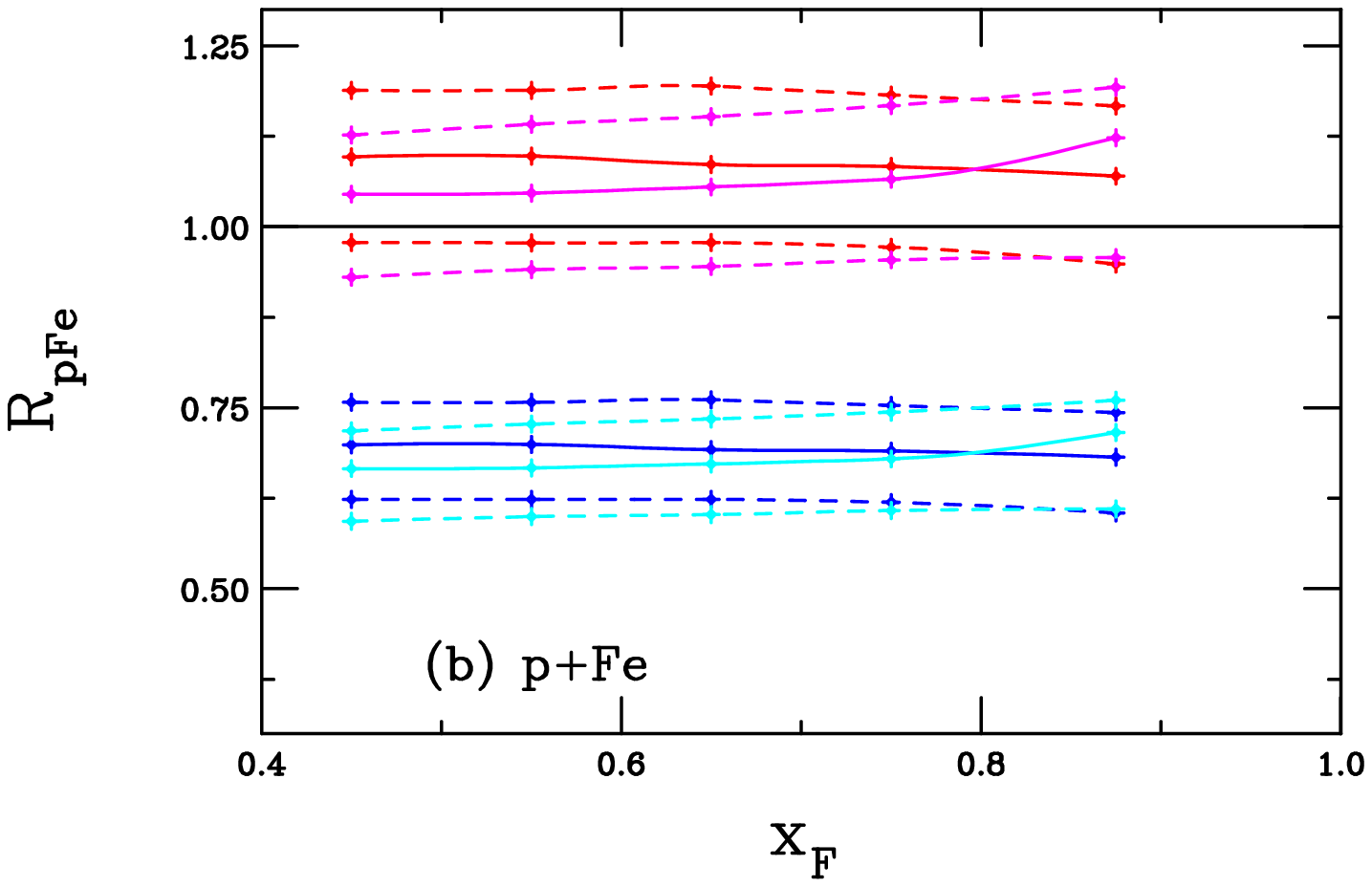}    
    \includegraphics[width=0.495\textwidth]{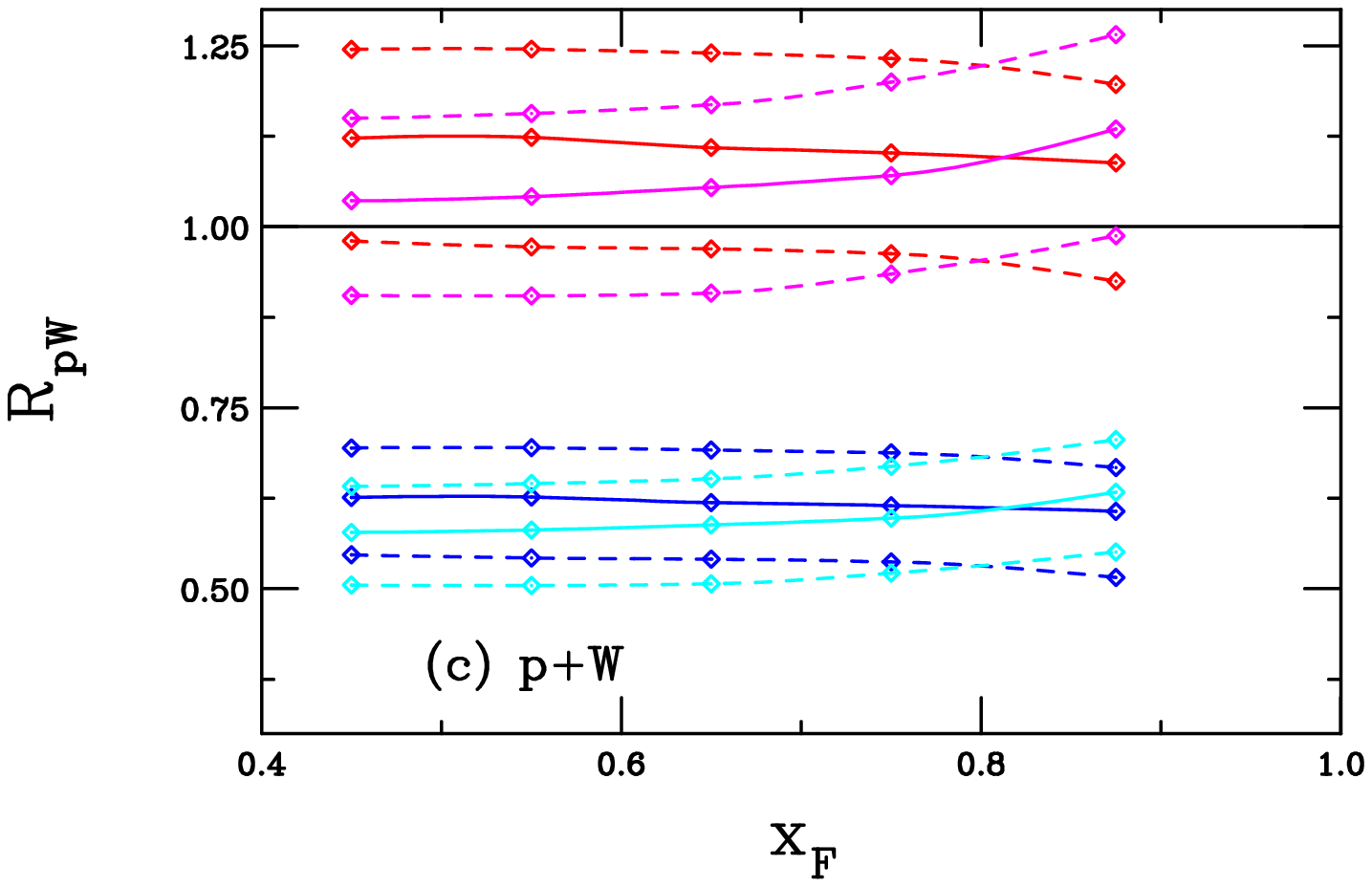}    
  \end{center}
  \caption[]{(Color online) The nuclear modification factors for $J/\psi$
    production in SeaQuest as a function of $x_F$ for pQCD production alone
    for carbon (a), iron (b) and tungsten (c) targets relative to deuterium.
    Results are shown for nPDF effects alone (red), nPDFs with an
    additional $k_T$ kick
    (magenta), nPDFs and absorption (blue), and nPDFs, absorption and
    $k_T$ broadening (cyan).  The solid lines shown the results with the
    central EPPS16 set while the dashed curves denote the limits of adding
    the EPPS16 uncertainties in quadrature.
  }
\label{shad_xF}
\end{figure}

\begin{figure}
  \begin{center}
    \includegraphics[width=0.495\textwidth]{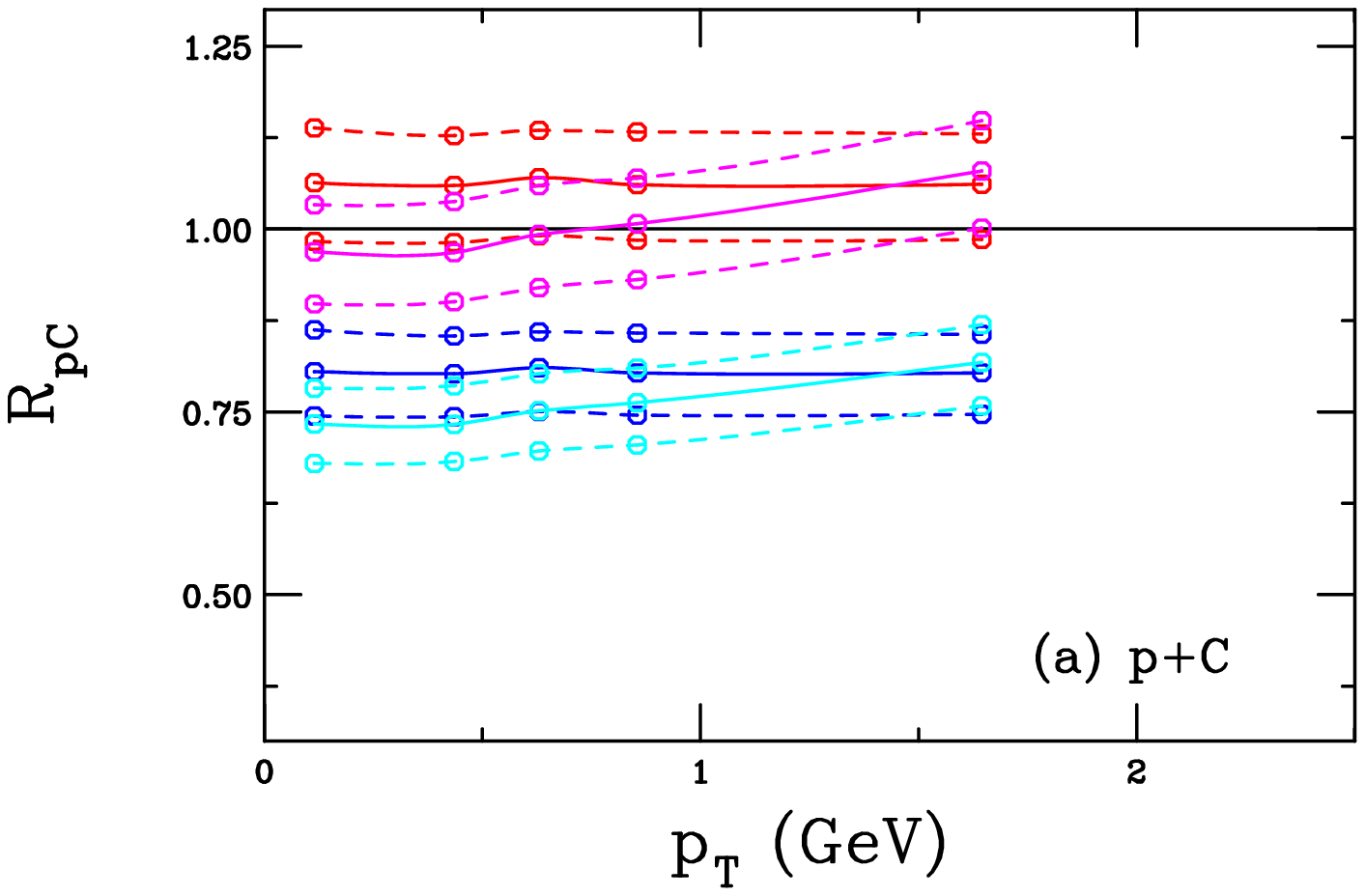}    
    \includegraphics[width=0.495\textwidth]{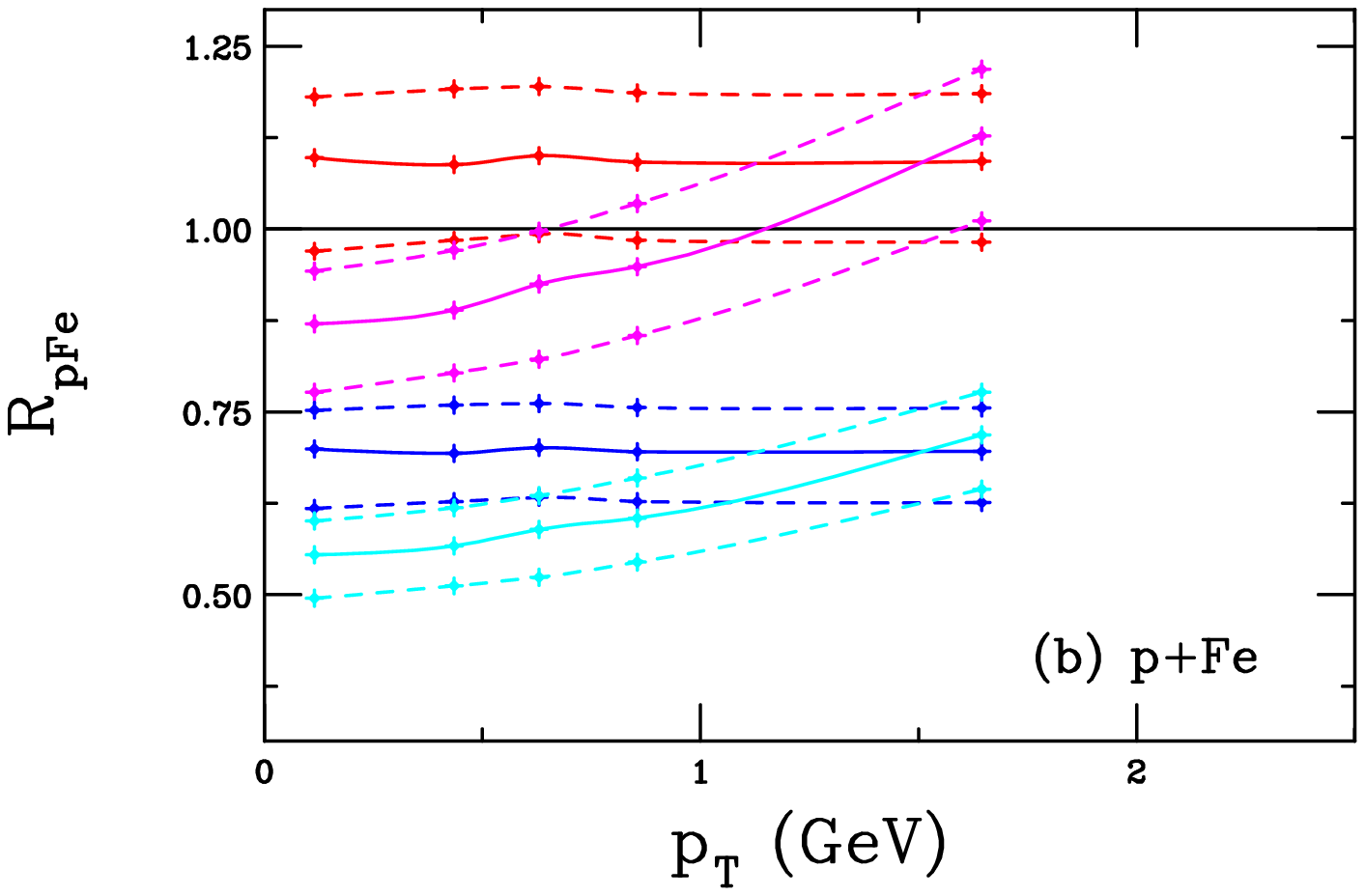}    
    \includegraphics[width=0.495\textwidth]{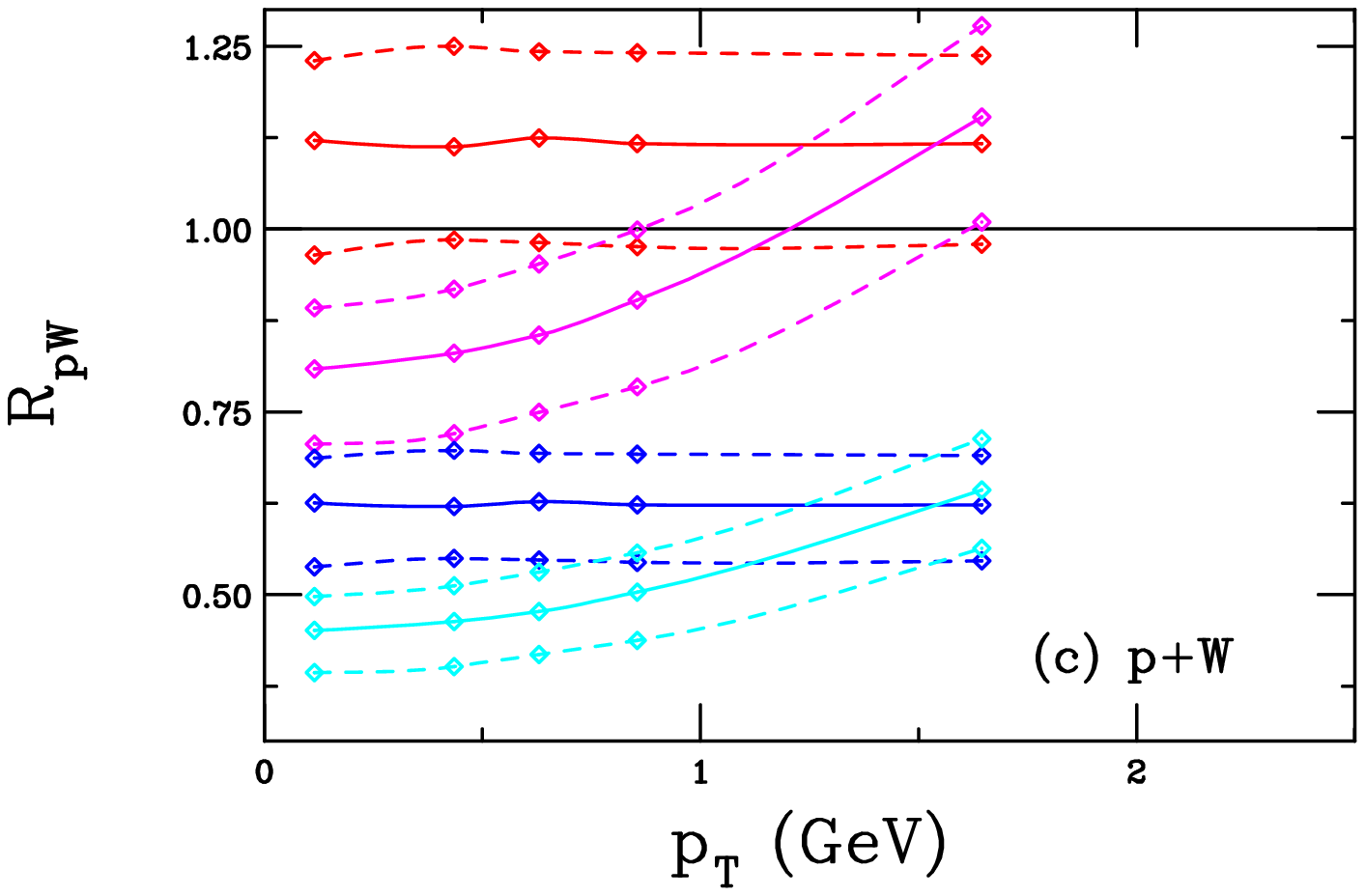}    
  \end{center}
  \caption[]{(Color online) The nuclear modification factors for $J/\psi$
    production in SeaQuest as a function of $p_T$ for pQCD production alone
    for carbon (a), iron (b) and tungsten (c) targets relative to deuterium.
    Results are shown for EPPS16 nPDFs alone (red), EPPS16 with an additional
    $k_T$ kick
    (magenta), nPDFs with absorption (blue), and nPDFs, absorption and
    $k_T$ broadening (cyan).  The solid lines show the results with the
    central EPPS16 set while the dashed curves denote the limits of adding
    the EPPS16 uncertainities in quadrature.
  }
\label{shad_pT}
\end{figure}

The focus of the discussion in this section is on only the nuclear effects
described in Sec.~\ref{pQCD}.  In this case, the cross sections employed in
the calculation of $R_{pA}$ are
\be
\sigma_{pA} = \sigma_{\rm CEM}(pA) = S_A^{\rm abs} F_C \sum_{i,j} 
\int_{4m^2}^{4m_H^2} ds
\int dx_1 \, dx_2~ F_i^p(x_1,\mu_F^2,k_T)~ F_j^A(x_2,\mu_F^2,k_T)~ 
\hat\sigma_{ij}(\hat{s},\mu_F^2, \mu_R^2) \, \, , 
\label{sigCEM_pA}
\ee
where
\be
F_j^A(x_2,\mu_F^2,k_T) & = & R_j(x_2,\mu_F^2,A) f_j(x_2,\mu_F^2) G_A(k_T) \, \, \\
F_i^p(x_1,\mu_F^2,k_T) & = & f_i(x_1,\mu_F^2) G_p(k_T) \, \, .
\ee
In the case of the deuteron target, $R_j \equiv 1$ in EPPS16, and
$\delta k_T^2 = 0$
in Eq.~(\ref{delkt2}) so that $g_A(k_T) = g_p(k_T)$.  It is also assumed that
$\sigma_{\rm abs} = 0$ since the proton and neutron in the deuteron are far
separated so the incident proton is likely to interact with only one of the
two nucleons in the deuteron.  Thus $\sigma_{p {\rm d}} = 2\sigma_{\rm CEM}(pp)$.
As was the case in Eq.~(\ref{FSgauss}), here the broadening is applied as
\begin{eqnarray}
  g_A(k_T) = G_p(k_{T_1}) G_A(k_{T_2}) \, \, ,
\end{eqnarray}
discussed in Sec.~\ref{kTkick}.
Results are shown in Fig.~\ref{shad_xF} as a function of $x_F$ and in
Fig.~\ref{shad_pT} as a function of $p_T$
for effects on the three nuclear
targets.  

The effects are added sequentially in these figures.  Only nPDF
effects are common to all the calculated ratios.  These nPDF results are
presented with the central, best fit, set given by the solid curves while the
uncertainties added in quadrature are outlined by the dashed curves.
First nPDF effects alone
are shown by the red points and curves.  In this case, $g_A(k_T) = g_p(k_T)$, no
additional broadening is taken into account due to the nuclear target.
The magenta points and curves show the effect of enhanced $k_T$ broadening in
the nuclear target.  Next, absorption is added in the two cases with
nPDF effects plus absorption shown in the blue points and curves while all three
effects: nPDF, absorption and enhanced $k_T$ broadening are shown by the
cyan points and curves.

The results with nPDF effects
alone are generally independent of $x_F$ and $p_T$
and all show $R_{pA} > 1$.  As seen in Fig.~\ref{shad_ratios}, the $x_F$ and
$p_T$ acceptance of the SeaQuest experiment restricts the $x_2$ range in the
nuclear targets to a rather narrow window close to $x_2 \sim 0.1$ where
antishadowing is dominant.  The low scale of the calculation, not much greater
than the minimum scale of EPPS16, indicates that the uncertainties will be
larger than at higher scales.  The EPPS16 nPDF uncertainties are large.
The effects grow in magnitude as the target mass is increased
from $A = 12$ to $A = 184$, as do the widths of the uncertainty bands.  It is
worth noting that only the lower limit of EPPS16 band gives $R_{pA} < 1$
in this case.

The effect of $k_T$ broadening in the nucleus relative to the deuteron changes
the shape of the $p_T$ distribution, particularly at low $p_T$.  By adding the
$\delta k_T^2$ due to multiple scattering to the intrinsic $k_T$ needed for
$J/\psi$ production in $p+p$ collisions, the average $p_T$ is increased in
$p+A$ collisions. The tails of the $p_T$ distributions are typically less
affected, particularly at collider energies where the enhanced $k_T$
broadening in a nuclear has to be large to produce a substantial effect because
the added $\delta k_T^2$ is small compared to $\langle p_T \rangle$ of the
$J/\psi$. In addition, if $\langle k_T^2 \rangle_A^{1/2}$ is much less than the
mass of the particle to which it is applied the effect is reduced.  For
example, the effect of $k_T$ broadening, even in $p+p$, is much smaller for
bottom quarks than for charm quarks, see Refs.~\cite{RVazi1,RVazi2}.

Because the mass scale of the $J/\psi$ is larger than the applied $k_T$ kick,
the low center of mass energy of the SeaQuest experiment results in a larger
effect than at collider energies and even than at other fixed-target energies.
This effect is particularly visible on the $p_T$ dependence of the suppression
factor shown in Fig.~\ref{shad_pT}.  The broadening results in a substantial
increase in $R_{pA}$ with $p_T$, especially for heavier targets where
$\delta k_T^2$ is a factor of 2-3 larger than the change in carbon.

It is notable, however, that there is also a slight change in the $x_F$
dependence of $R_{pA}$ due to broadening since
$x_F = 2m_T \sinh y/\sqrt{s_{NN}}$.  This change is significantly smaller
than as a
function of $p_T$ because $x_F$ depends instead on $m_T = \sqrt{m^2 + p_T^2}$
and $m$ and $p_T$ are of comparable magnitude over the measured $p_T$ range.
Finally, there would be no change due to $k_T$ broadening if the
results were given as a function of rapidity instead.

Rather strong absorption is required to nullify the effects of antishadowing
and produce $R_{pA} < 1$, as seen when absorption is added to the calculation.
Since a constant 9~mb cross section has been assumed, as inferred for
midrapidity ($x_F \sim 0$) in Ref.~\cite{LWV}, absorption does not change the
dependence of $R_{pA}$ on $x_F$ and $p_T$.
The effective exponent $\alpha$ obtained with $\sigma_{\rm abs} = 9$~mb is 0.88.

Reference~\cite{LWV} also showed
that $\sigma_{\rm abs}$ could increase significantly with $x_F$ for
$x_F > 0.25$, resulting in a decrease of $R_{pA}$ with $x_F$.
In that reference and in other works, see for example
Ref.~\cite{GM,E772_eloss,Arleo,RV_e866}, the decrease with $x_F$
was associated with
energy loss, either in the initial state or the final state.  Here it is
argued that this behavior could be attributed to intrinsic charm.  This
possibility will be tested in the next section where the effects on an intrinsic
charm contribution is studied.

In Ref.~\cite{LWV}, away from midrapidity but for
$x_F \leq 0.25$, the effective absorption cross section could be interpreted as
starting to decrease with $x_F$,
especially for higher fixed-target energies.  This could
be attributed, for example, to the $J/\psi$ being formed outside the target.
Given the forward acceptance of the SeaQuest experiment, well forward 
of the acceptances of most of the experiments included in Ref.~\cite{LWV}, it is
possible that a lower value of $\sigma_{\rm abs}$ could be extracted from
comparison with the SeaQuest data.

\subsection{$R_{pA}$ Including Intrinsic Charm}
\label{RpA:pQCD+IC}

Intrinsic charm is added to the calculations of $R_{pA}$ here.  In this case,
\be
\sigma_{pA} & = & \sigma_{\rm CEM}(pA) + \sigma_{\rm ic}^{J/\psi}(pA)
\label{sig_pA_sum} \\
\sigma_{p{\rm d}} & = & 2\sigma_{\rm CEM}(pp) + \sigma_{\rm ic}^{J/\psi}(pA) \, \, .
\label{sig_pd_sum}
\ee
Here $\sigma_{\rm CEM}(pA)$ was defined in Eq.~(\ref{sigCEM_pA}) while
$\sigma_{\rm ic}^{J/\psi}(pA)$ is given in Eq.~(\ref{icsigJpsi_pA}).
Note that, unlike the nuclear volume-like mass dependence of the perturbative
QCD effects, the nuclear surface-like $A$ dependence is assumed to apply to
the deuteron.  Later in this section, this assumption will be relaxed to check
the dependence of the results on this assumption.
Three values of $P_{{\rm ic}\, 5}^0$ in
Eq.~(\ref{icdenom}) are shown in the following
four figures: 0.1\%, 0.31\% and 1\%.  This range
can be taken as an uncertainty band on intrinsic charm.

\begin{figure}
  \begin{center}
    \includegraphics[width=0.495\textwidth]{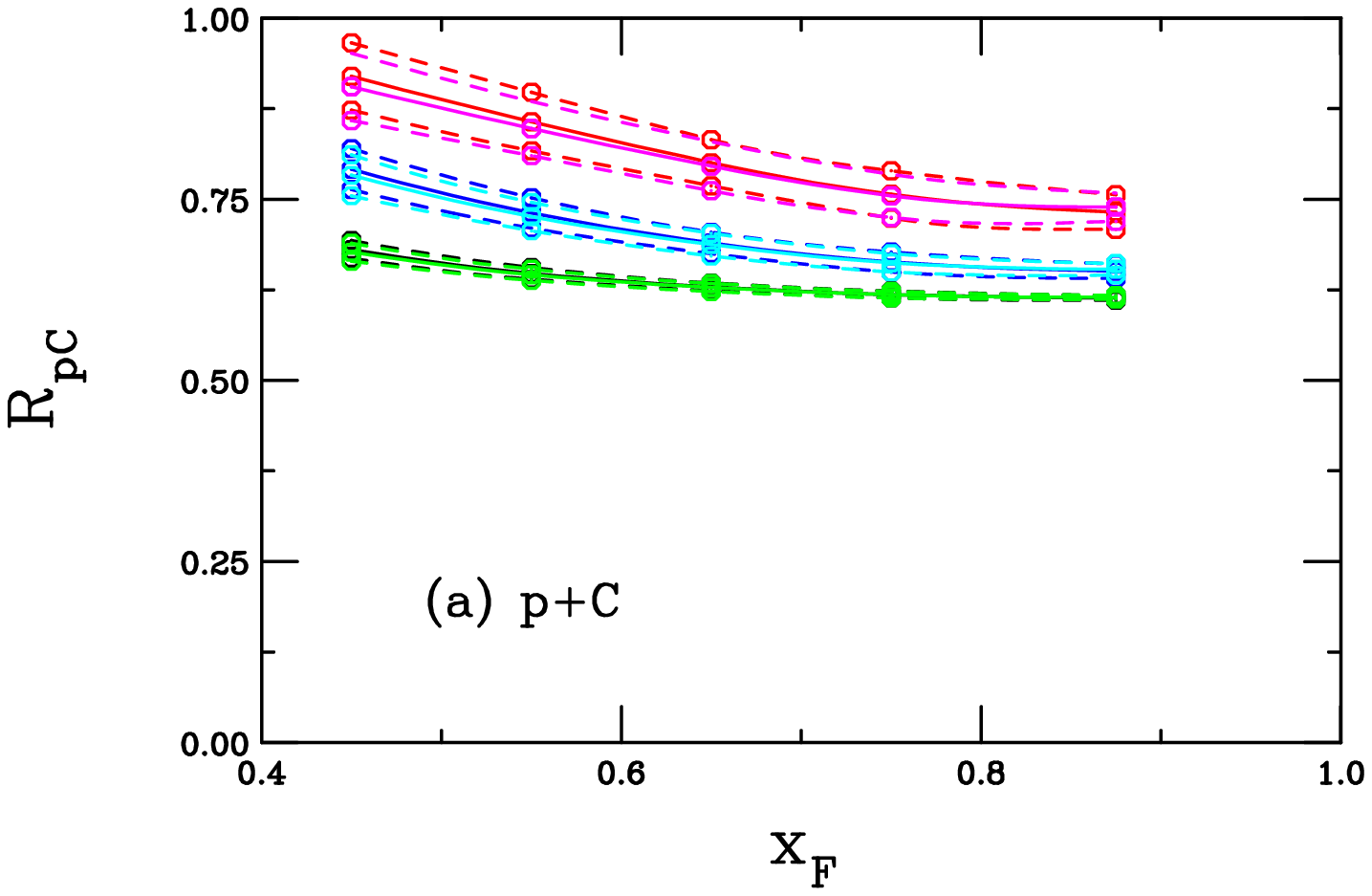}    
    \includegraphics[width=0.495\textwidth]{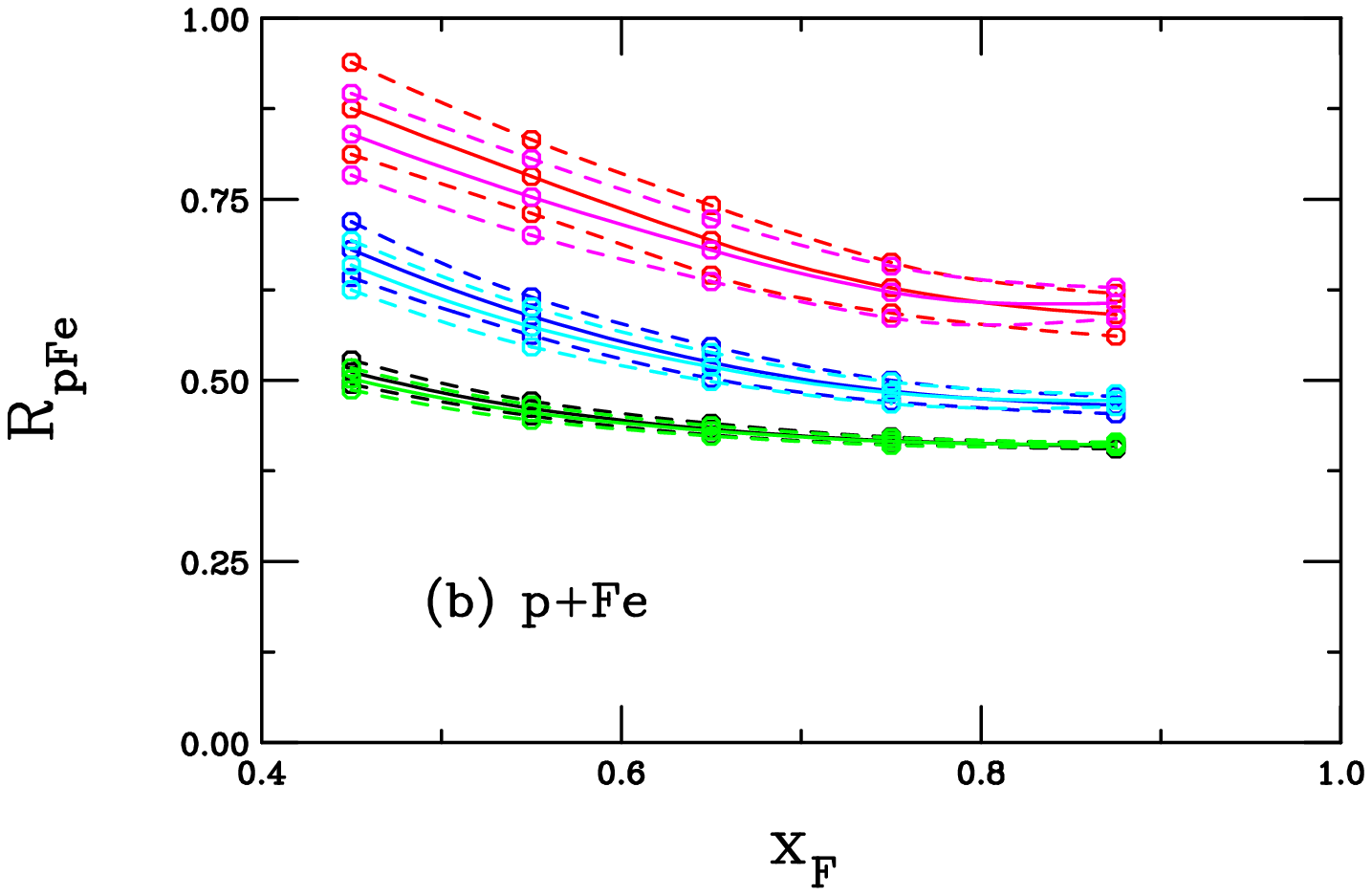}    
    \includegraphics[width=0.495\textwidth]{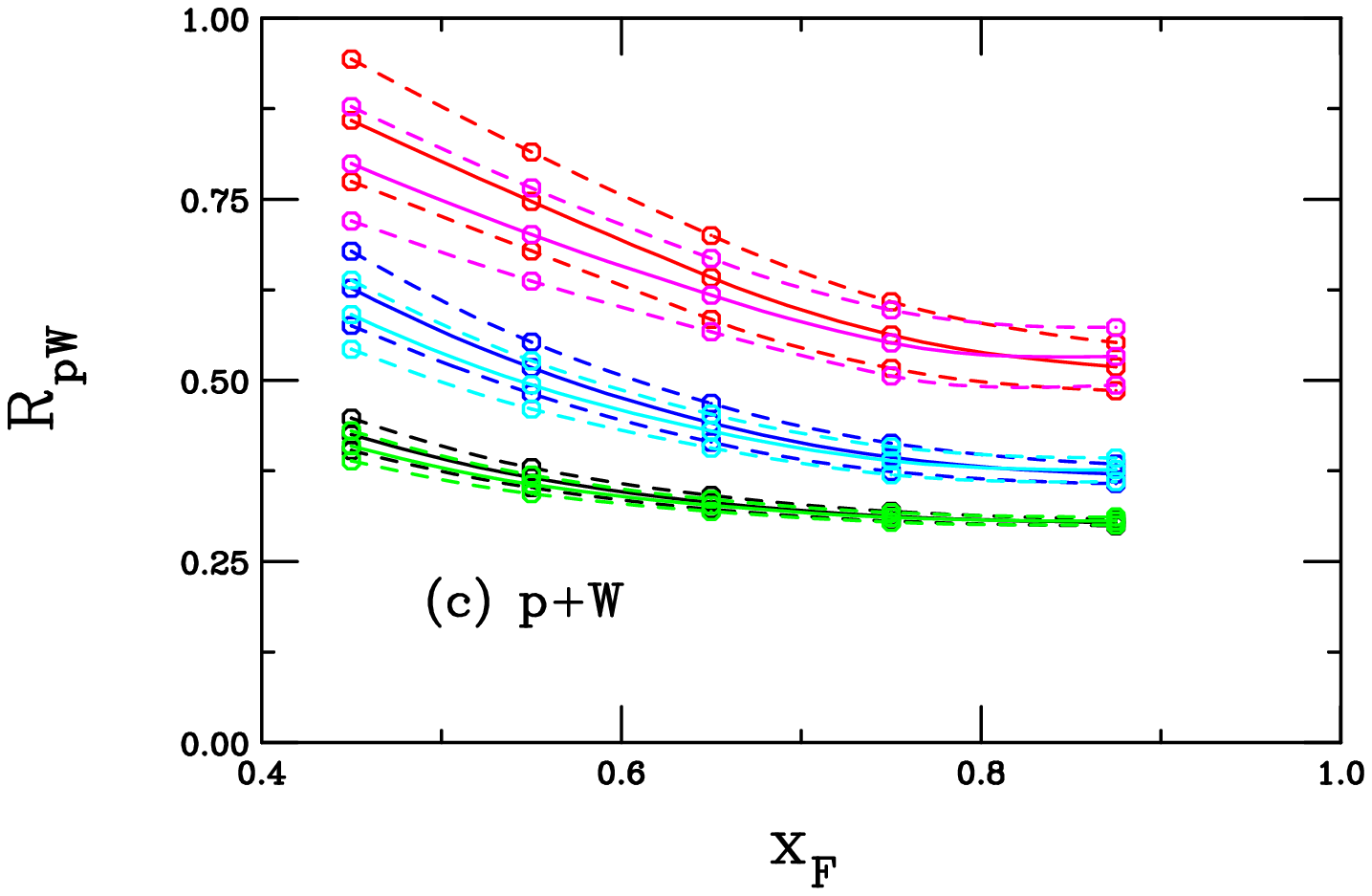}    
  \end{center}
  \caption[]{(Color online) The nuclear modification factors for $J/\psi$
    production in SeaQuest as a function of $x_F$ for the combined pQCD and
    intrinsic charm cross section ratios
    for carbon (a), iron (b) and tungsten (c) targets relative to deuterium.
    All calculations are shown with $\sigma_{\rm abs} = 0$.
    Results with nPDF effects and the same $k_T$ in $p+{\rm d}$
    and $p+A$ are shown in
    the red, blue and black curves while nPDF effects with an enhanced $k_T$
    kick in the nucleus are shown in the magenta, cyan and green curves.
    The probability for IC production is 0.1\% in the red and magenta curves;
    0.31\% in the blue and cyan curves; and 1\% in the black and green curves.
    The solid lines show the results with the
    central EPPS16 set while the dashed curves denote the limits of adding
    the EPPS16 uncertainities in quadrature.
  }
\label{IC_ratios_noabs_xF}
\end{figure}

\begin{figure}
  \begin{center}
    \includegraphics[width=0.495\textwidth]{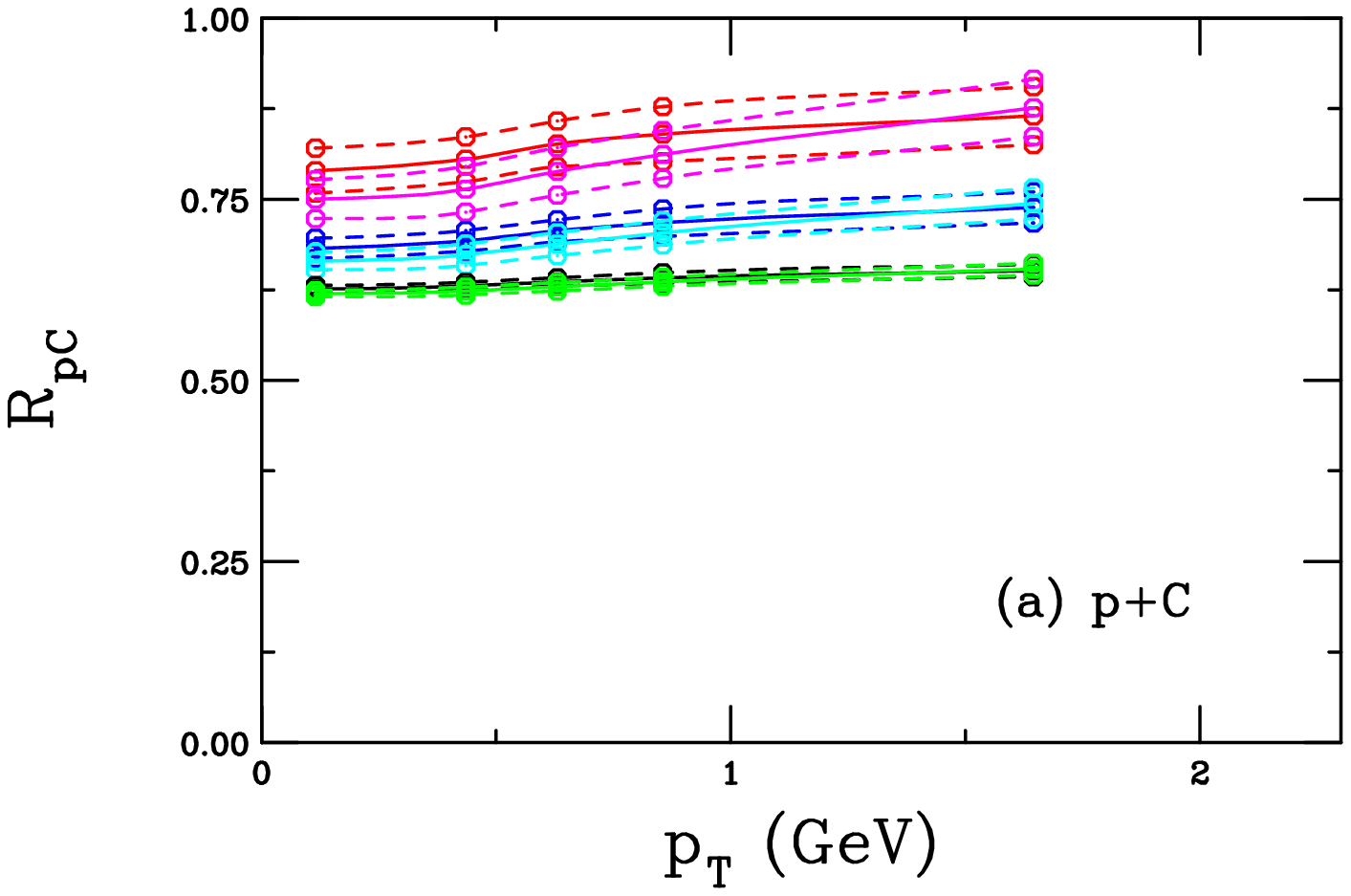}    
    \includegraphics[width=0.495\textwidth]{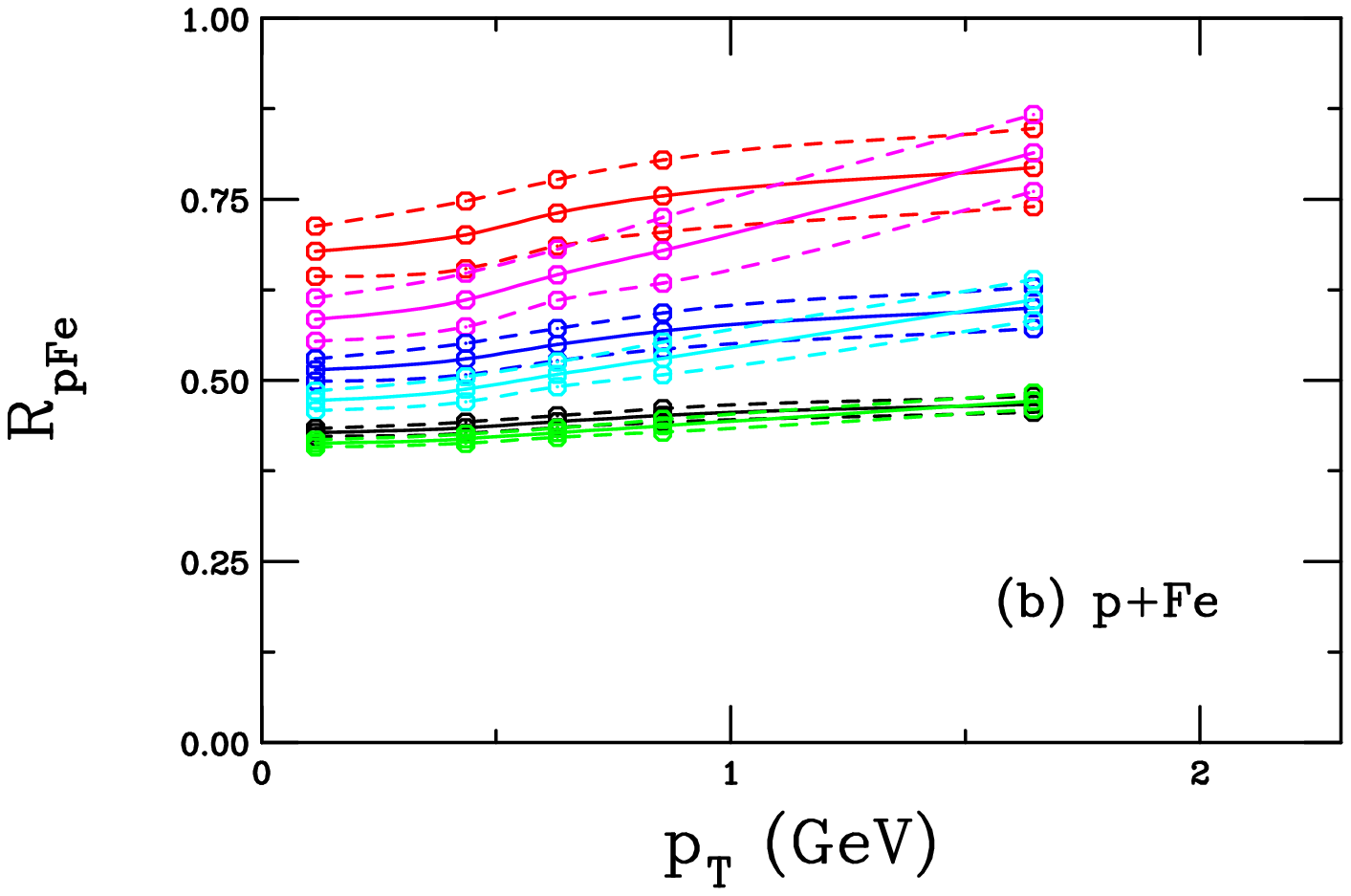}    
    \includegraphics[width=0.495\textwidth]{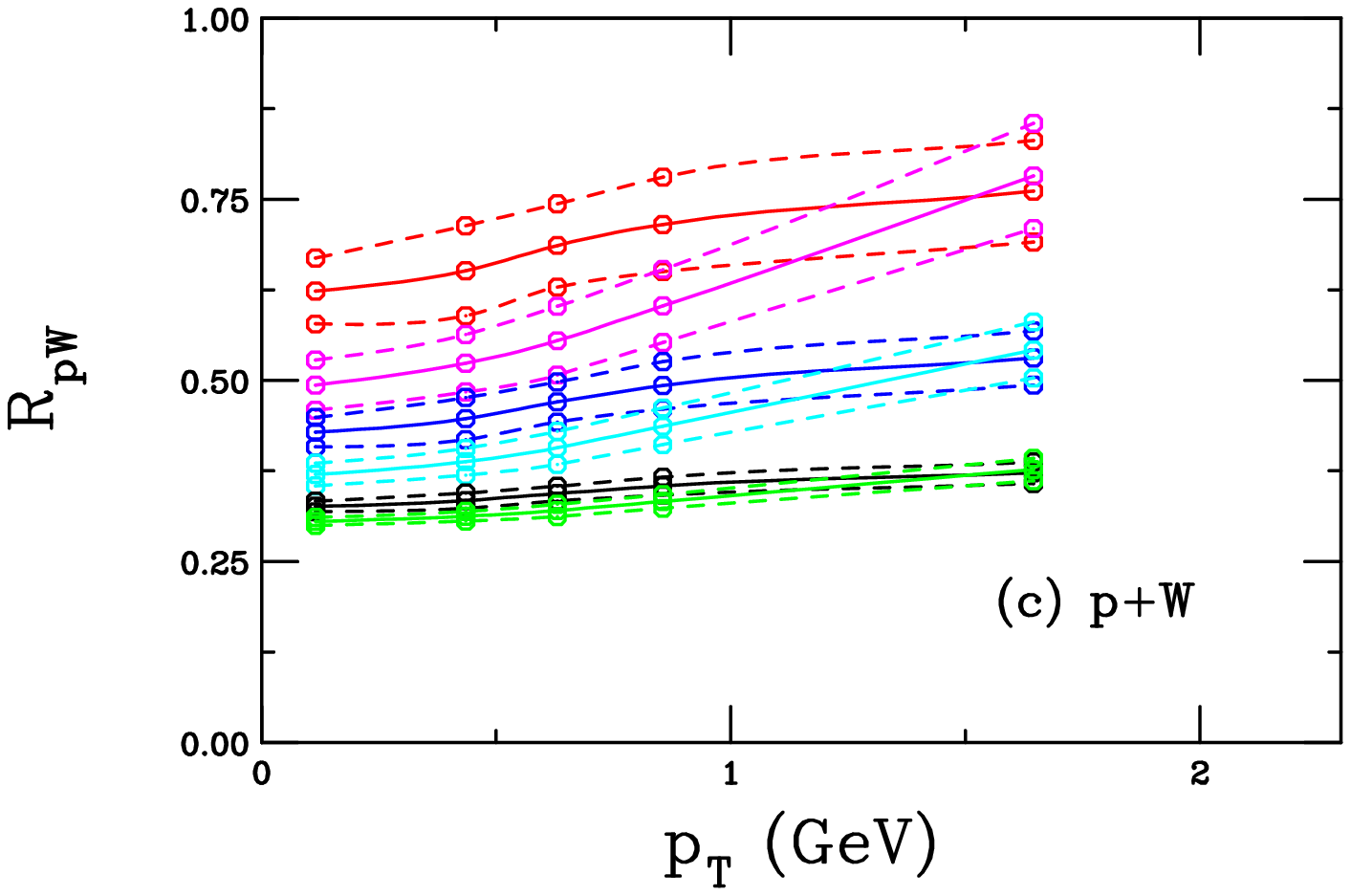}    
  \end{center}
  \caption[]{(Color online) The nuclear modification factors for $J/\psi$
    production in SeaQuest as a function of $p_T$ for the combined pQCD and
    intrinsic charm cross section ratios
    for carbon (a), iron (b) and tungsten (c) targets relative to deuterium.
    All calculations are shown with $\sigma_{\rm abs} = 0$.
    Results with nPDF effects and the same $k_T$ in $p+{\rm d}$
    and $p+A$ are shown in
    the red, blue and black curves while nPDFs with an enhanced $k_T$
    kick in the nucleus are shown in the magenta, cyan and green curves.
    The probability for IC production is 0.1\% in the red and magenta curves;
    0.31\% in the blue and cyan curves; and 1\% in the black and green curves.
    The solid lines shown the results with the
    central EPPS16 set while the dashed curves denote the limits of adding
    the EPPS16 uncertainities in quadrature.
  }
\label{IC_ratios_noabs_pT}
\end{figure}

\begin{figure}
  \begin{center}
    \includegraphics[width=0.495\textwidth]{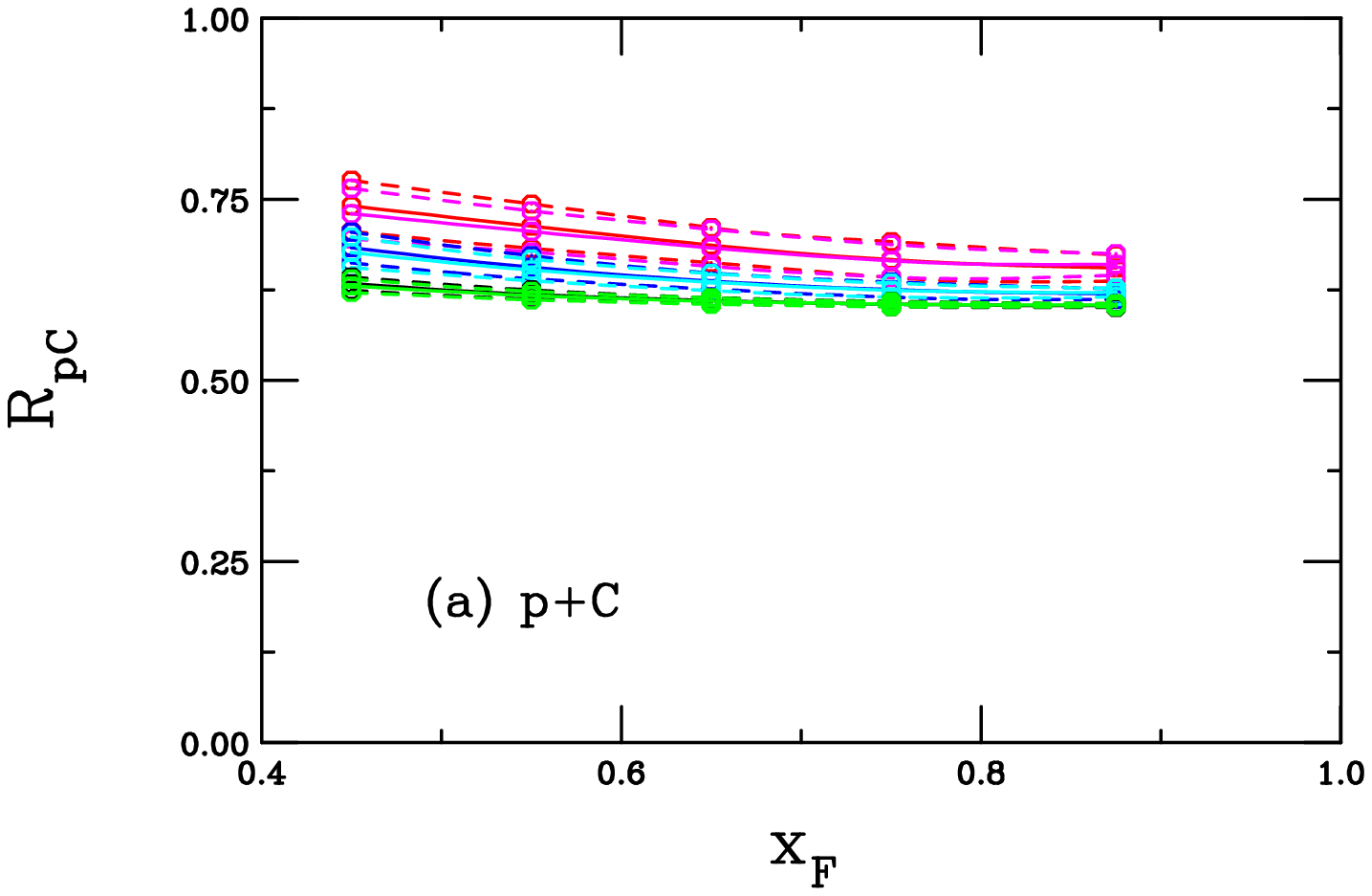}    
    \includegraphics[width=0.495\textwidth]{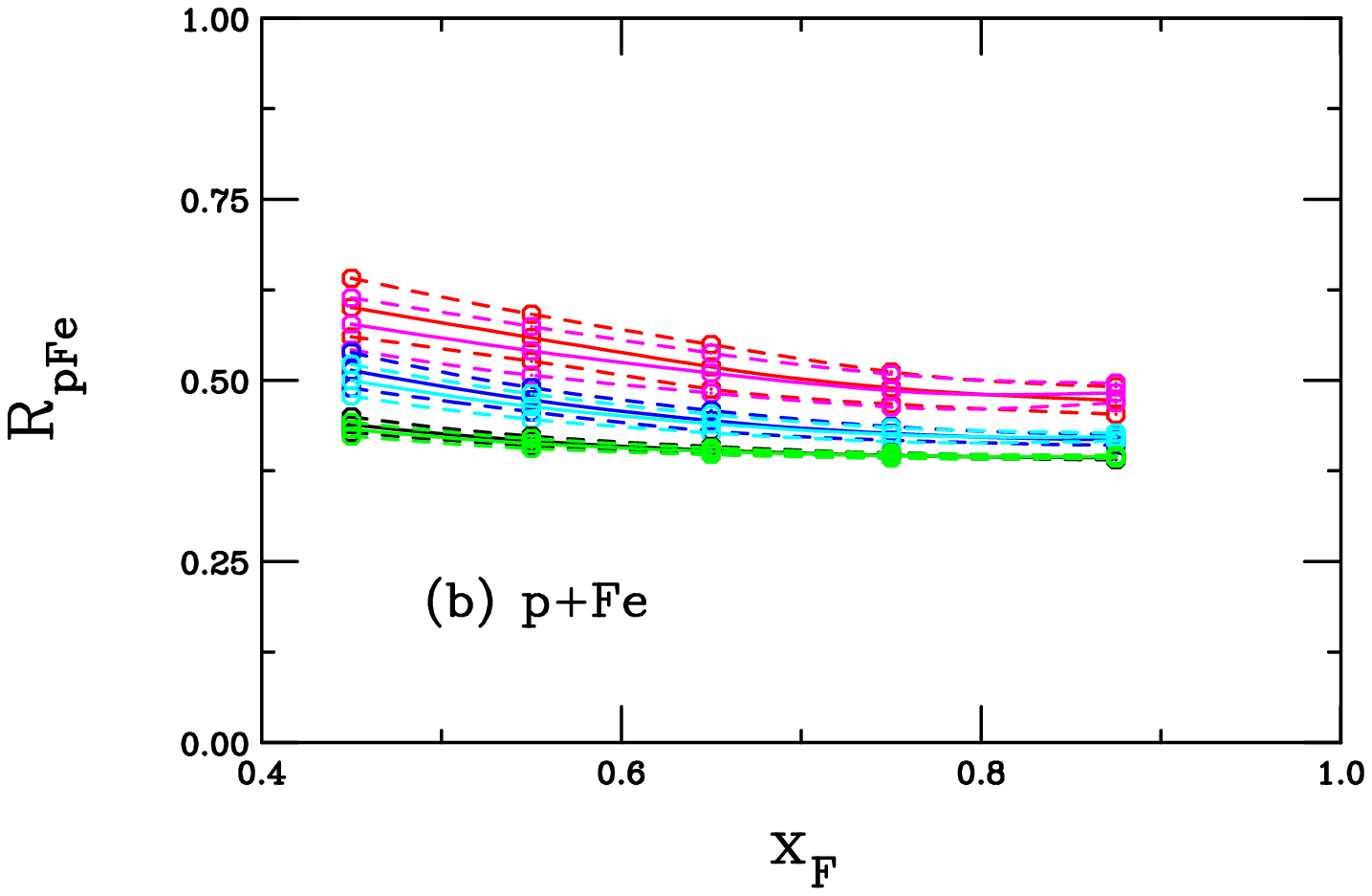}    
    \includegraphics[width=0.495\textwidth]{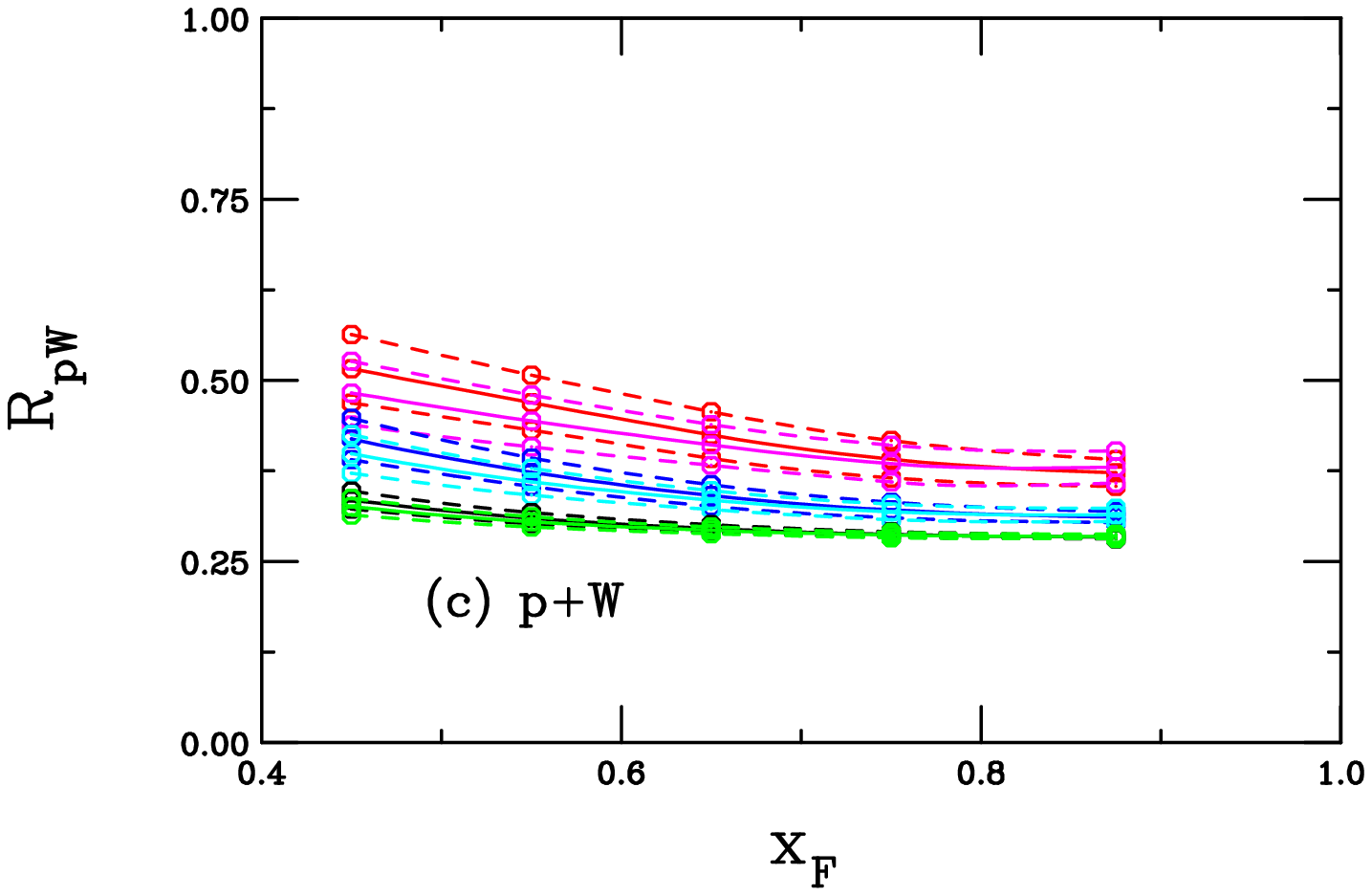}    
  \end{center}
  \caption[]{(Color online) The nuclear modification factors for $J/\psi$
    production in SeaQuest as a function of $x_F$ for the combined pQCD and
    intrinsic charm cross section ratios
    for carbon (a), iron (b) and tungsten (c) targets relative to deuterium.
    All calculations are shown with nuclear absorption included.
    Results with EPPS16 and the same $k_T$ in $p+{\rm d}$
    and $p+A$ are shown in
    the red, blue and black curves while EPPS16 with an enhanced $k_T$
    kick in the nucleus are shown in the magenta, cyan and green curves.
    The probability for IC production is 0.1\% in the red and magenta curves;
    0.31\% in the blue and cyan curves; and 1\% in the black and green curves.
    The solid lines shown the results with the
    central EPPS16 set while the dashed curves denote the limits of adding
    the EPPS16 uncertainities in quadrature.
\label{IC_ratios_abs_xF}}
\end{figure}

\begin{figure}
  \begin{center}
    \includegraphics[width=0.495\textwidth]{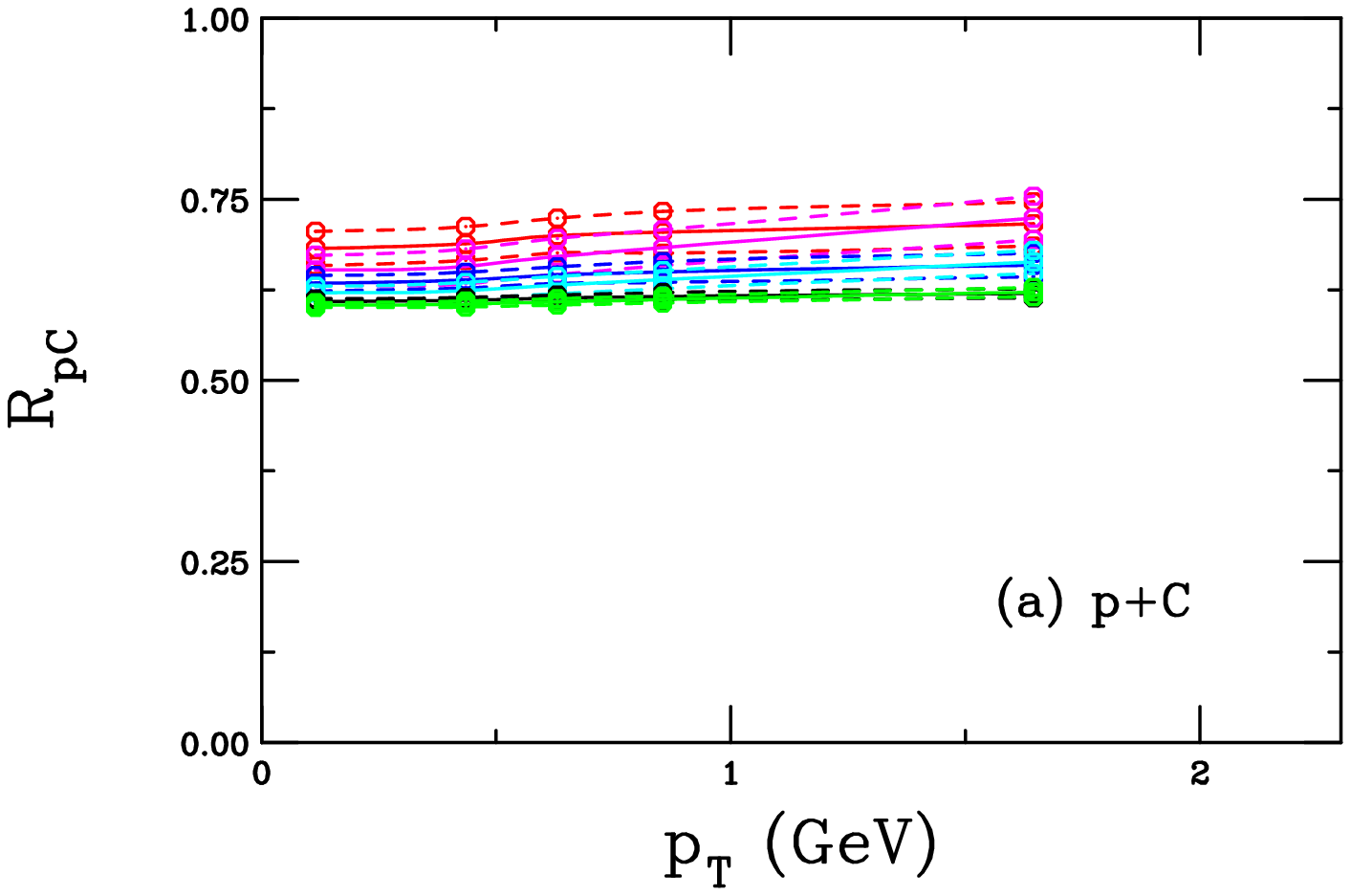}    
    \includegraphics[width=0.495\textwidth]{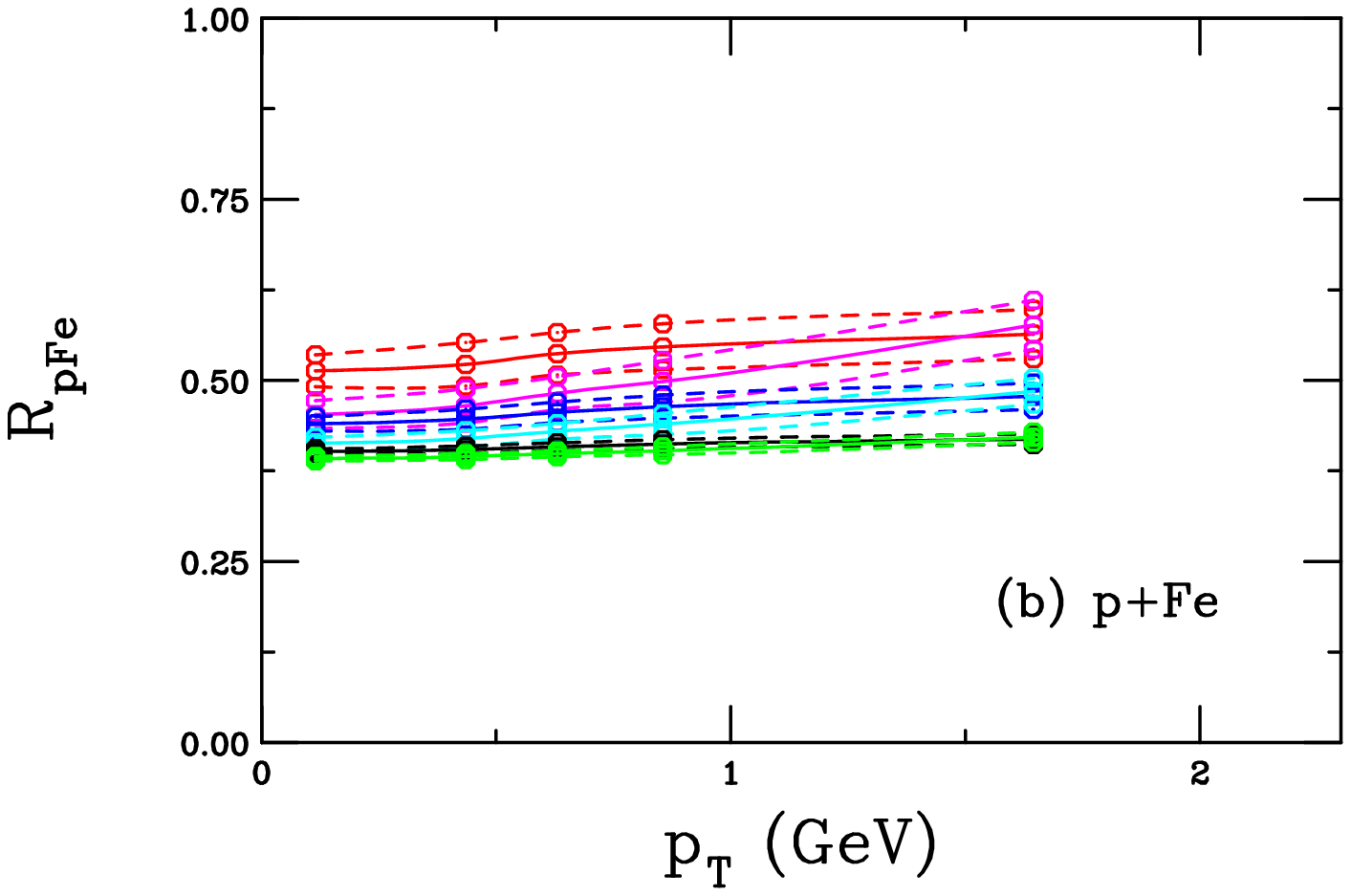}    
    \includegraphics[width=0.495\textwidth]{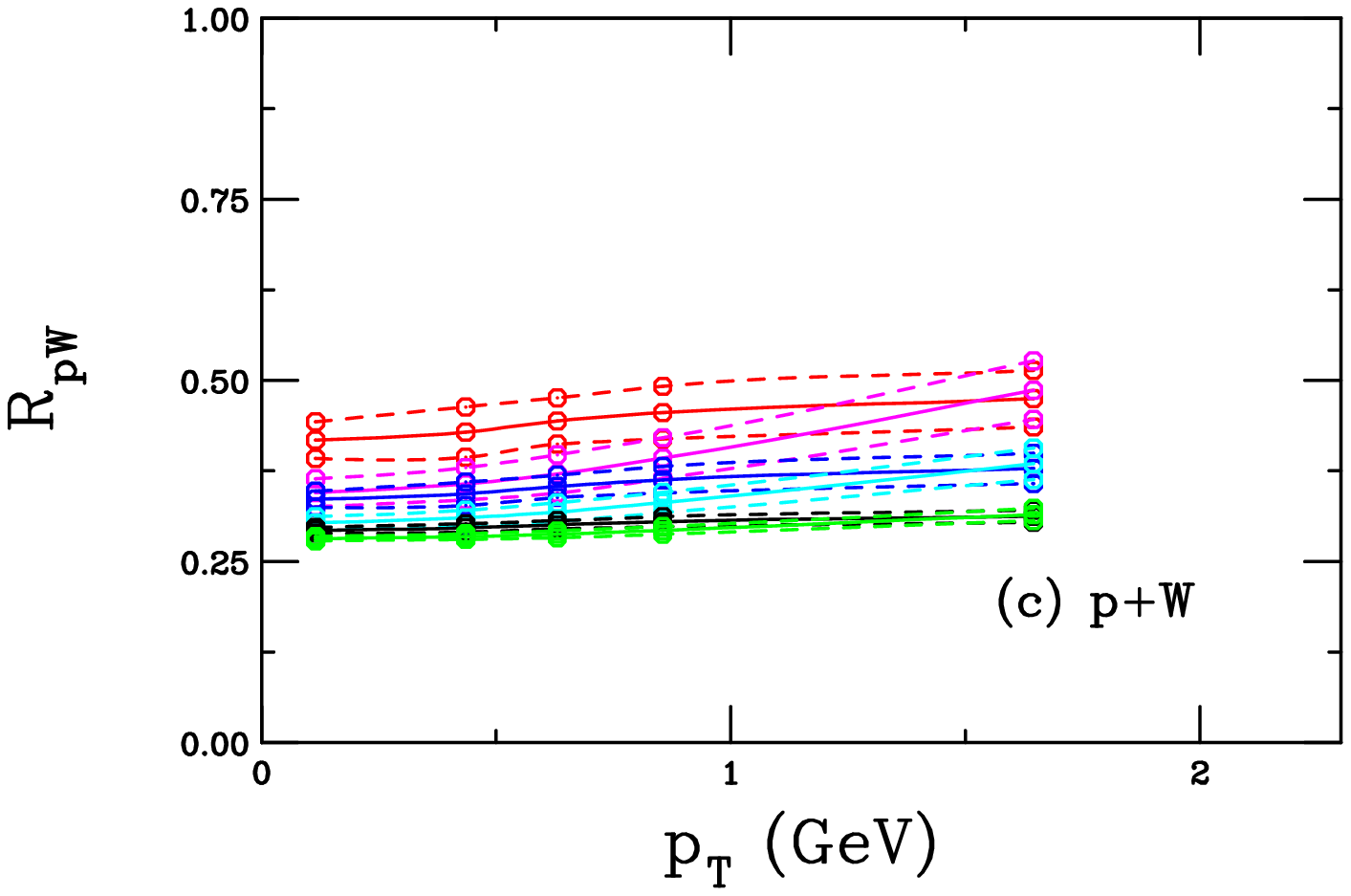}    
  \end{center}
  \caption[]{(Color online) The nuclear modification factors for $J/\psi$
    production in SeaQuest as a function of $pT$ for the combined pQCD and
    intrinsic charm cross section ratios
    for carbon (a), iron (b) and tungsten (c) targets relative to deuterium.
    All calculations are shown with nuclear absorption included.
    Results with EPPS16 and the same $k_T$ in $p+{\rm d}$
    and $p+A$ are shown in
    the red, blue and black curves while EPPS16 with an enhanced $k_T$
    kick in the nucleus are shown in the magenta, cyan and green curves.
    The probability for IC production is 0.1\% in the red and magenta curves;
    0.31\% in the blue and cyan curves; and 1\% in the black and green curves.
    The solid lines shown the results with the
    central EPPS16 set while the dashed curves denote the limits of adding
    the EPPS16 uncertainities in quadrature.
\label{IC_ratios_abs_pT}}
\end{figure}

As in the previous section, the effects are added sequentially in these
figures.  Now nPDF effects and intrinsic charm are
common to all the calculated ratios.  The EPPS16 results are
presented with the central, best fit, set given by the solid curves while the
uncertainties added in quadrature are outlined by the dashed curves.
Those calculations, affecting $\sigma_{\rm CEM}$ are added to the intrinsic
charm cross section, as in Eqs.~(\ref{sig_pA_sum}) and (\ref{sig_pd_sum}).
Results are shown
without absorption, $\sigma_{\rm abs}=0$, in Figs.~\ref{IC_ratios_noabs_xF} and
\ref{IC_ratios_noabs_pT} while absorption is included with
$\sigma_{\rm abs} = 9$~mb in
Figs.~\ref{IC_ratios_abs_xF} and \ref{IC_ratios_abs_pT}.  

In these figures, the red, blue and black curves show the nPDF effects
with $P_{{\rm ic}\, 5}^0 = 0.1$\%, 0.31\% and 1\% respectively in
Figs.~\ref{IC_ratios_noabs_xF} and
\ref{IC_ratios_noabs_pT}.  The solid curves show
the EPPS16 central value while the dashed curves outline the uncertainty band.
For these calculations, $g_A(k_T) = g_p(k_T)$.  The magenta, cyan and green
solid and dashed curves include enhanced $k_T$ broadening in
the nuclear target for $P_{{\rm ic}\, 5}^0 = 0.1$\%, 0.31\% and 1\% respectively.
The same color scheme is used in Figs.~\ref{IC_ratios_abs_xF} and
\ref{IC_ratios_abs_pT} but with absorption included in $\sigma_{\rm CEM}(pA)$.

The nuclear dependence of intrinsic charm is not shown separately because no
other nuclear effects are active with this component, only the $A^\beta$
dependence shown in Eq.~(\ref{icsigJpsi_pA}).  Thus $R_{pA}$ for intrinsic
charm alone would be independent of $x_F$ and $p_T$.  Since intrinsic charm
is included only in the initial proton and the $J/\psi$ comes on shell without
a hard interaction with the target, there are no nPDF effects.  For the same
reason, the intrinsic charm
contribution is unaffected by multiple scattering of the proton in the target
nucleus.

The effect of the intrinsic charm contribution is immediately clear in
Fig.~\ref{IC_ratios_noabs_xF}: $R_{pA} < 1$ without nuclear absorption and is
now a decreasing function of $x_F$.  It is also clear that, at least for the
SeaQuest energy and in its forward acceptance, the intrinsic charm cross
section is comparable to the perturbative QCD cross section.  As
$P_{{\rm ic}\, 5}^0$
increases from 0.1\% to 1\%, the intrinsic charm contribution comes to
dominate the nuclear dependence as a function of $x_F$.
The most immediate qualitative measure of this
dominance is apparent in the width of the EPPS16 uncertainty band.
These bands, are wide in Figs.~\ref{shad_xF} and \ref{shad_pT}
without intrinsic charm,
narrower but still relatively far apart for $P_{{\rm ic}\, 5}^0 = 0.1$\%, and
barely distinguishable for $P_{{\rm ic}\, 5}^0 = 1$\% because, in this case,
the nuclear dependence is wholly dominated by intrinsic charm.

There is some notable separation between the different assumed values of
$P_{{\rm ic}\, 5}^0$ without absorption included.  The difference due to the $k_T$
broadening, apparent for the red and magenta curves in
Fig.~\ref{IC_ratios_noabs_xF}, especially for the iron and tungsten targets,
is almost indistinguishable for $P_{\rm ic}^0 = 1$\%, in the black
and green curves.  The dominance of the intrinsic charm contribution as
$P_{{\rm ic}\, 5}^0$ increases also becomes obvious in the weakening of the $x_F$
dependence with increasing $P_{{\rm ic}\, 5}^0$.
When a 1\% probability is assumed,
the results are almost independent of $x_F$ except in the lowest $x_F$ bins,
as would be expected if $J/\psi$ production was via the intrinsic charm
contribution alone.

Similar behavior is observed as a function of $p_T$ in
Fig.~\ref{IC_ratios_noabs_pT}.  Even without an enhanced $k_T$ broadening in
the nucleus, a rise in $R_{pA}$ with $p_T$ is still observed, although it is not
as strong a function of $p_T$ without broadening as it is with it.
In this case,
with the same $k_T$ kick in $p+A$ as in $p+p$, there is an increase compared to
$p_T \rightarrow 0$ which then levels off in the last two, larger $p_T$, bins.
On the other hand, when broadening is included, the rise in $R_{pA}$ with $p_T$
continues over the entire $p_T$ range shown.  The perturbative QCD and intrinsic
charm contributions are thus competitive with each other as long as
$P_{{\rm ic}\,5}^0$ is low, 0.31\% or less, but not for $P_{{\rm ic}\,5}^0=1$\%
where the distinction
between calculations with and without broadening is negligible.

Results including absorption are given in Figs.~\ref{IC_ratios_abs_xF} and
\ref{IC_ratios_abs_pT}.  The strong absorption cross section
employed in these calculations, $\sigma_{\rm abs} = 9$~mb,
results in dominance of the intrinsic charm
contribution even when $P_{{\rm ic}\, 5}^0$ is as low as 0.1\%.
Note, however, that
the dependence of $R_{pA}$ with $x_F$ and $p_T$ is not significantly affected
by an assumed 1\% probability for intrinsic charm because, for this value of
$P_{{\rm ic}\, 5}^0$, the intrinsic charm contribution is already dominant with
$\sigma_{\rm abs} = 0$, the higher probability overcomes the stronger surface
nuclear target
dependence of intrinsic charm.  Now, with absorption included, the $A$
dependence of the two contributions are similar in magnitude since
$S_A^{\rm abs} \approx A^\alpha = A^{0.88}$, see the discussion under
Eq.~(\ref{sigfull}) in Sec.~\ref{absorption}.  Even though $\alpha$ is still
greater than $\beta = 0.71$, assumed for intrinsic charm \cite{NA3,VBH1},
the two values are more comparable.

The calculations including absorption further compresses the pQCD uncertainties
due to nPDF effects and $k_T$ broadening.
The differences between the calculations
with nPDF effects only and nPDFs with $k_T$ broadening on the perturbative part
are almost indistinguishable as a function of $x_F$ except for the heaviest
targets, as seen in Fig.~\ref{IC_ratios_abs_xF}.  The same can be seen as a
function of $p_T$ in Fig.~\ref{IC_ratios_abs_pT}.  In this figure, the
differences between the results without and with $k_T$ broadening are only
visible for $P_{{\rm ic}\, 5}^0 = 0.1$\%.

\begin{figure}
  \begin{center}
    \includegraphics[width=0.495\textwidth]{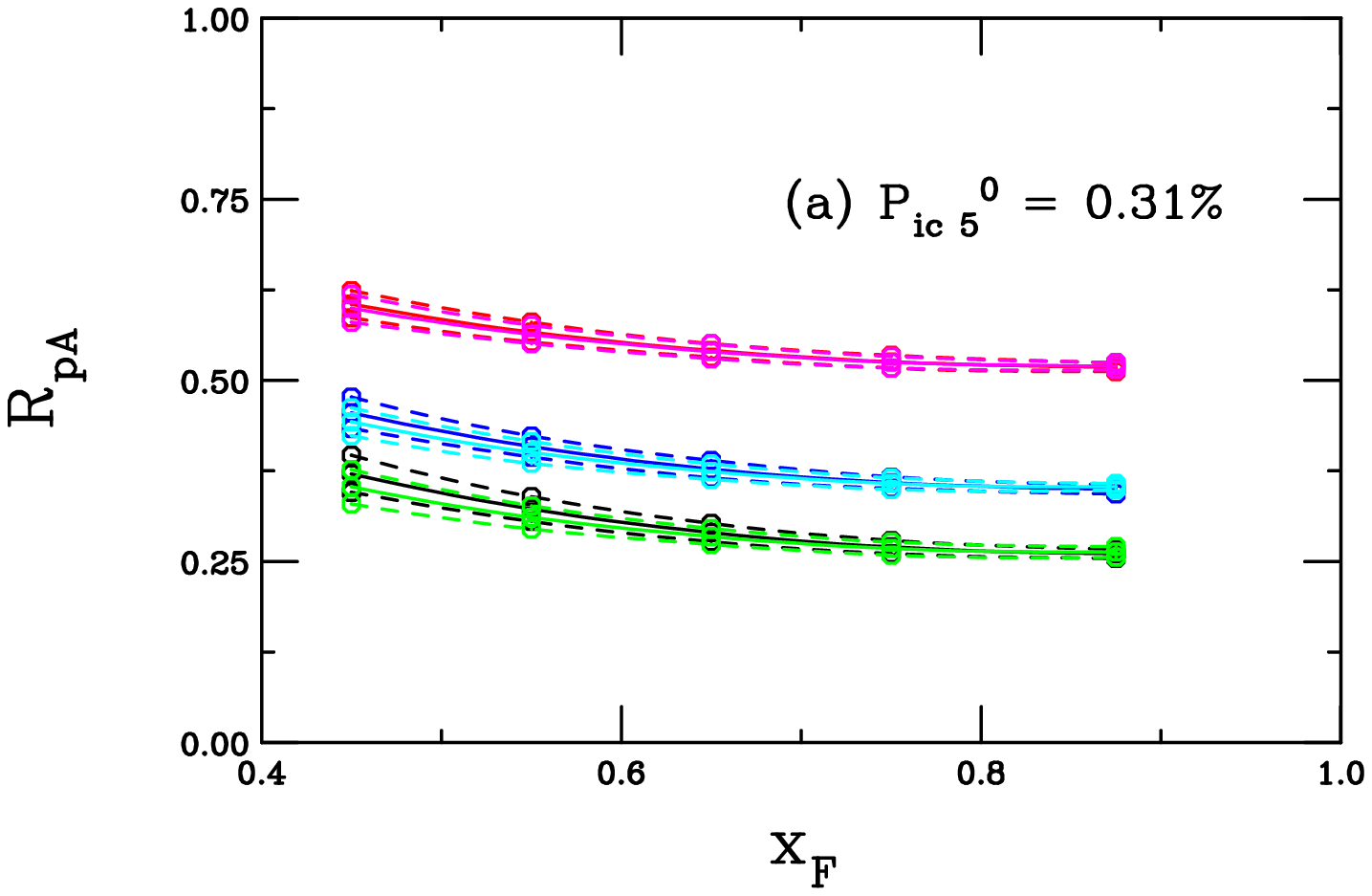}    
    \includegraphics[width=0.495\textwidth]{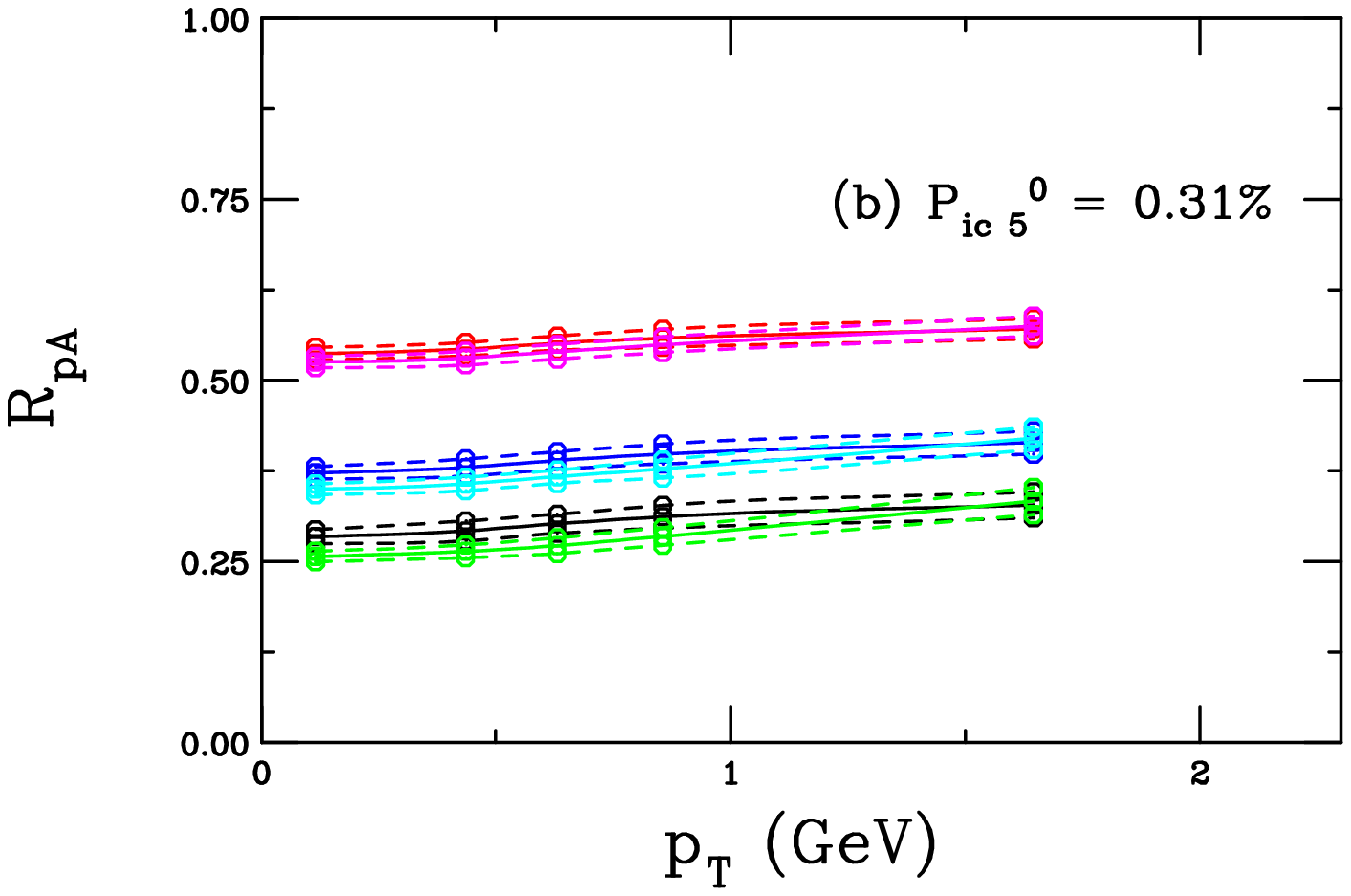}    
  \end{center}
  \caption[]{(Color online) The nuclear modification factors for $J/\psi$
    production in SeaQuest as a function of $x_F$ (a) and $p_T$ (b) for the
    combined pQCD and intrinsic charm cross section ratios
    for carbon (red and magenta curves), iron (blue and cyan curves) and
    tungsten (black and green curves) targets relative to a proton target.
    All calculations are shown with nuclear absorption included.
    Results with EPPS16 and the same $k_T$ in $p+p$ and $p+A$ are shown in
    the red, blue and black curves while EPPS16 with an enhanced $k_T$
    kick in the nucleus are shown in the magenta, cyan and green curves.
    The probability for IC production is 0.31\% in all cases.
    The solid lines shown the results with the
    central EPPS16 set while the dashed curves denote the limits of adding
    the EPPS16 uncertainities in quadrature.
\label{IC_ratios_pp}}
\end{figure}

\begin{figure}
  \begin{center}
    \includegraphics[width=0.495\textwidth]{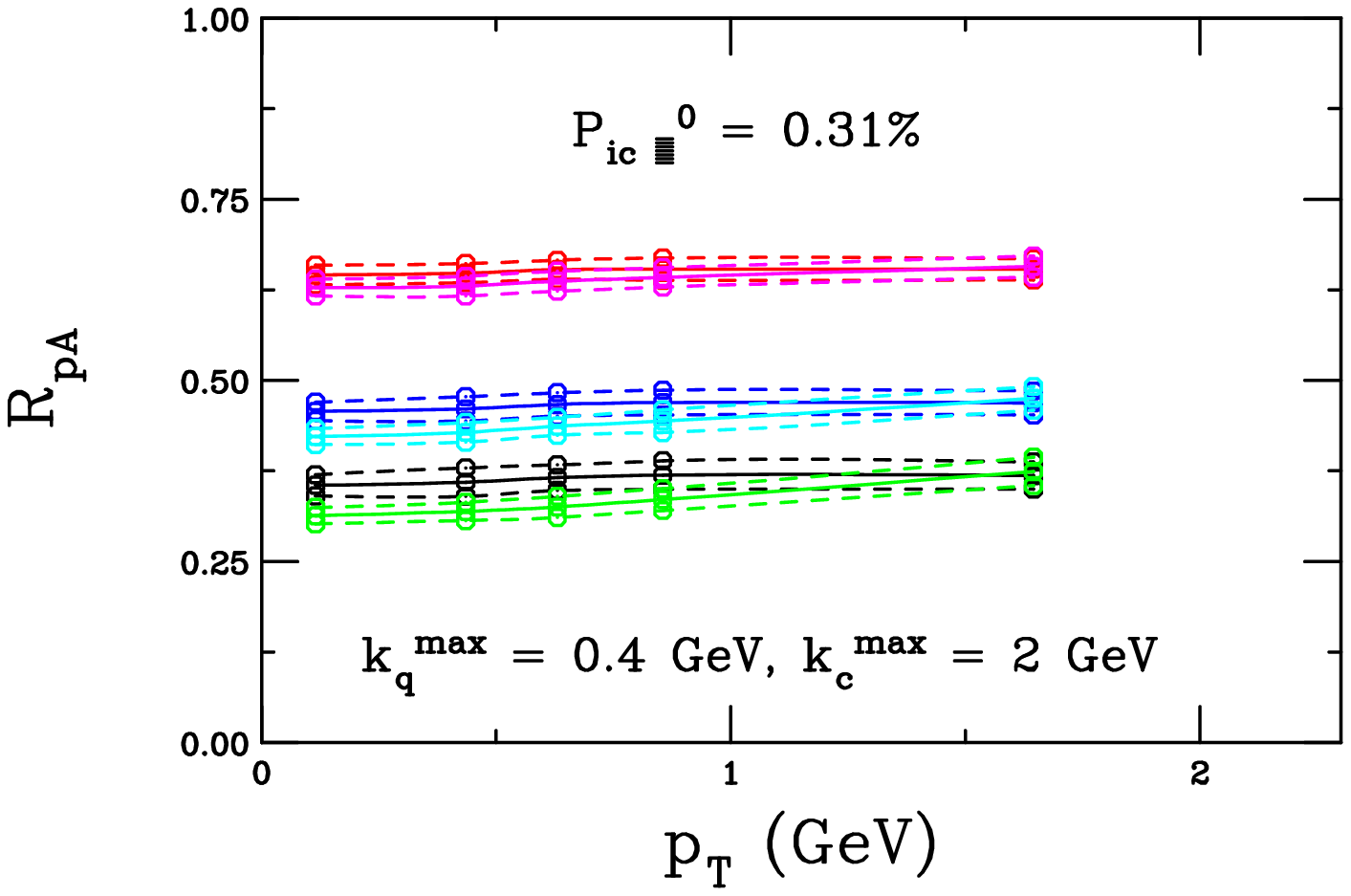}    
  \end{center}
  \caption[]{(Color online) The nuclear modification factors for $J/\psi$
    production in SeaQuest as a function of $p_T$ for the
    combined pQCD and intrinsic charm cross section ratios
    for carbon (red and magenta curves), iron (blue and cyan curves) and
    tungsten (black and green curves) targets relative to a proton target.
    All calculations are shown with nuclear absorption included.
    Results with EPPS16 and the same $k_T$ in $p+{\rm d}$
    and $p+A$ are shown in
    the red, blue and black curves while EPPS16 with an enhanced $k_T$
    kick in the nucleus are shown in the magenta, cyan and green curves.
    The probability for IC production is 0.31\% in all cases.  A broader IC
    distribution is assumed here, corresponding to the blue dashed curve in
    Fig.~\protect\ref{ic_dists}(b).
    The solid lines shown the results with the
    central EPPS16 set while the dashed curves denote the limits of adding
    the EPPS16 uncertainities in quadrature.
\label{IC_ratios_2kt}}
\end{figure}

The last part of this section further tests the assumptions made about how
intrinsic charm is implemented.  First, the assumption regarding whether one
treats the deuterium target as a nucleus or like a proton is tested, with
results shown in Fig.~\ref{IC_ratios_pp}.  Next, the influence of the range of
$k_T$ integration on the intrinsic charm $p_T$ distribution is checked in
Fig.~\ref{IC_ratios_2kt}.

When calculating $R_{pA}$ with a deuteron target, it has
so far been assumed that the surface-like $A$ dependence also applies to the
deuteron, unlike the assumption that $\sigma_{\rm abs} = 0$, a pure volume-like
$A$ dependence, for perturbative QCD effects.  This has the effect of reducing
the intrinsic charm contribution in $p+{\rm d}$
interactions relative to $p+A$ for
heavier nuclei.  In Fig.~\ref{IC_ratios_pp}, it is assumed that, also in the
case of intrinsic charm, the deutron target is effectively the same as a proton
target, with $\beta = 1$ in Eq.~(\ref{icsigJpsi_pA}).  Instead of showing
all values of $P_{{\rm ic}\, 5}^0$, only calculations with
$P_{{\rm ic}\, 5}^0 = 0.31$\% are presented.

Results are given for all three targets with EPPS16 and
absorption.  Calculations both with and without an intrinsic $k_T$ enhancement
are presented.  The $R_{pA}$ ratios are, from top to bottom, $p+{\rm C}$
(red without $k_T$ enhancement and magenta with); $p+{\rm Fe}$
(blue and cyan); and $p+{\rm W}$ (black and green).  Making the
intrinsic charm contribution in $p+{\rm d}$ the same
as in $p+p$ has the effect of reducing the magnitude of the ratios by a few
percent without changing the shape of $R_{pA}$ as a function of $x_F$ or $p_T$.
The difference would be slightly larger for a smaller value of
$P_{{\rm ic}\, 5}^0$ but would not change the overall result significantly.  Thus,
at least from the point of view of statistics with a proton or deuteron target,
there is no reason, calculationally, to prefer one over the other.

To end this section, Fig.~\ref{IC_ratios_2kt} shows how
$R_{pA}$ could be modified as
a function of $p_T$ if a wider range of $k_T$ integration for intrinsic charm
had been chosen.  In particular, $k_q^{\rm max} = 0.4$~GeV and
$k_c^{\rm max} = 2$~GeV are assumed since this choice resulted in a broader
$J/\psi$ distribution from intrinsic charm with a correspondingly reduced peak
of the $p_T$ distribution at low $p_T$, in the range of the SeaQuest acceptance.
Results are not also shown for the lower values because, although the $p_T$
distribution becomes somewhat narrower, the effect is not large enough to
modify the results.  The nuclear suppression factor
is only shown as a function of $p_T$ 
because the range of $k_T$ integration has no effect on the $x_F$
distribution of the produced $J/\psi$, see Fig.~\ref{ic_dists}.

Similar to the results shown in Fig.~\ref{IC_ratios_pp}, only
$P_{{\rm ic}\, 5}^0 = 0.31$\% is presented for all three targets, without and with
enhanced $k_T$ broadening, in addition to nPDF effects and absorption.  The
broader intrinsic charm $p_T$ distribution does not change the overall
suppression of $R_{pA}$ significantly but it does show a larger difference
between calculations with additional $k_T$ broadening and those without.

\section{Conclusions}
\label{conclusions}

The low center of mass energy of the SeaQuest experiment, as well as its
forward acceptance, make it an ideal environment for probing the existence of
an intrinsic
charm contribution to $J/\psi$ production.  The center of mass energy is 
a factor of 3.1 above the $J/\psi$ production threshold.  This relatively
low energy makes the perturbative QCD cross section compatible with or less than
the $J/\psi$ cross section from intrinsic charm, depending on the experimental
$x_F$ range.  Higher center of mass energies increase the $J/\psi$ cross
section in the CEM
dramatically, see, {\it e.g.} Ref.~\cite{NVF}, while the intrinsic charm
contribution grows more slowly, depending only on $\sigma_{pN}^{\rm in}$, see
Eq.~(\ref{icsign}).  In addition, the high $x_F$ range covered by
SeaQuest is exactly the region where intrinsic charm should dominate production.
As seen in a comparison of the $x_F$ distributions in Figs.~\ref{cem_pp}(a)
and \ref{ic_dists}(a), the CEM cross section decreases by an order of magnitude
over the SeaQuest $x_F$ acceptance, with a maximum at $x_F = 0$, outside the
SeaQuest acceptance.  On the other hand, the
peak of intrinsic charm probability distribution is at $x_F \approx 0.53$,
within the range of the SeaQuest measurement.

A comparison of the SeaQuest $J/\psi$ production data on its nuclear targets
could set limits on $\sigma_{\rm abs}$ in the
perturbative QCD contribution and $P_{{\rm ic}\, 5}^0$, the probability of
the intrinsic charm contribution in the proton.

{\bf Acknowledgments}
I would like to thank C. Aidala, A. Angerami, C. Ayuso and V. Cheung
for helpful discussions. This work was supported by the Office of Nuclear
Physics in the U.S. Department of Energy under Contract DE-AC52-07NA27344 and
the LLNL-LDRD Program under Project No. 21-LW-034.

\end{document}